\documentclass[manuscript,screen]{acmart}

\usepackage[linesnumbered,ruled,vlined]{algorithm2e}
\usepackage{amsmath} 
\usepackage{amssymb} 
\usepackage{amsthm}
\usepackage{graphicx} 
\usepackage{float}
\usepackage{multirow} 
\usepackage{tabularx}
\usepackage{xcolor}
\usepackage{hyperref}
\usepackage{dsfont}
\usepackage{tcolorbox}

\SetKwInput{KwIn}{Input} 
\SetKwInput{KwOut}{Output}
\SetKwInput{KwParam}{Parameter}
\usepackage{pifont}
\newcommand{\embed}[1]{E(#1)}

\definecolor{LightRed}{rgb}{1,0.3,0.3}

\AtBeginDocument{%
  }

\setcopyright{acmlicensed}
\copyrightyear{2018}
\acmYear{2018}
\acmDOI{XXXXXXX.XXXXXXX}

\acmConference[Conference acronym 'XX]{Make sure to enter the correct
  conference title from your rights confirmation emai}{June 03--05,
  2018}{Woodstock, NY}
\acmISBN{978-1-4503-XXXX-X/18/06}

\begin{document}

\title{EMPRA: Embedding Perturbation Rank Attack against Neural Ranking Models}

\author{Amin Bigdeli}
\orcid{0009-0003-8977-9312}
\affiliation{%
  \institution{University of Waterloo}
  \city{Waterloo}
  \state{ON}
  \country{Canada}
}
\email{abigdeli@uwaterloo.ca}

\author{Negar Arabzadeh}
\orcid{0000-0002-4411-7089}
\affiliation{%
  \institution{University of Waterloo}
  \city{Waterloo}
  \state{ON}
  \country{Canada}
}
\email{Narabzad@uwaterloo.ca}

\author{Ebrahim Bagheri}
\orcid{0000-0002-5148-6237}
\affiliation{%
  \institution{University of Toronto}
  \city{Toronto}
  \state{ON}
  \country{Canada}
}
\email{ebrahim.bagheri@utoronto.ca}

\author{Charles L. A. Clarke}
\orcid{0000-0001-8178-9194}
\affiliation{%
  \institution{University of Waterloo}
  \city{Waterloo}
  \state{ON}
  \country{Canada}
}
\email{claclark@gmail.com}

\renewcommand{\shortauthors}{Bigdeli et al.}

\begin{abstract}
Recent research has shown that neural information retrieval techniques may be susceptible to adversarial attacks. Adversarial attacks seek to manipulate the ranking of documents, with the intention of exposing users to targeted content. In this paper, we introduce the \textit{Embedding Perturbation Rank Attack} (\texttt{EMPRA}) method, a novel approach designed to perform adversarial attacks on black-box Neural Ranking Models (NRMs). 
\texttt{EMPRA} manipulates sentence-level embeddings, guiding them towards pertinent context related to the query while preserving semantic integrity. This process generates adversarial texts that seamlessly integrate with the original content and remain imperceptible to humans.
Our extensive evaluation conducted on the widely-used MS MARCO V1 passage collection demonstrate the effectiveness of \texttt{EMPRA} against a wide range of state-of-the-art baselines in promoting a specific set of target documents within a given ranked results. Specifically, \texttt{EMPRA} successfully achieves a re-ranking of almost 96\% of target documents originally ranked between 51-100 to rank within the top 10. Furthermore, \texttt{EMPRA} does not rely on surrogate models for generating adversarial documents, enhancing its robustness against various victim NRMs in realistic settings.

\end{abstract}




\maketitle

\section{Introduction}

Despite significant advancements in Neural Ranking Models (NRMs), recent research highlights vulnerabilities and a possible lack of resilience to adversarial attacks and perturbations, within both queries and documents \cite{penha2022evaluating,wu2022certified,wu2023prada}. These attacks are crafted to either elevate or diminish the ranking of a target document, thereby amplifying or reducing the likelihood of users encountering the information it contains. Consequently, the presence of such attacks and the fragility of neural information retrieval systems may negatively impact the integrity and dependability of the results.

In the early days of the web search, adversarial attacks might take the form of term spamming, wherein query-related terms were repetitively inserted into a target document to enhance its ranking in the retrieved results~\cite{imam2019survey,castillo2011adversarial,sasaki2005spam}. These attacks were undertaken with the aim of engaging in black-hat Search Engine Optimization (SEO), wherein specific documents are targeted and their content manipulated to secure higher rankings in search results. This manipulation sought to increase the visibility of the content, exposing it to a larger audience \cite{gyongyi2005web}. However, given their relative simplicity, such term spamming tactics were susceptible to detection by spam filters ~\cite{cormack2011efficient,zhou2009osd}.
In contrast, more recent research in this space has been inspired by broader work in adversarial attacks on deep neural networks, which are often designed to manipulate the classification outcomes of these models \cite{song2020adversarial, ebrahimi2017hotflip, zou2019reinforced, maimon2022universal}. Most notably, various authors have focused on assessing the resilience of neural-based ranking models against adversarial attacks, including word substitution rank attacks, trigger generation-based attacks and prompt-based attacks \cite{liu2022order,wang2022bert,chen2023towards,wu2023prada}. This work is based on the fact that neural ranking models learn the semantic mapping between the query and document during the training process. As such, adding/replacing terms and sentences that are semantically similar to the original text and are capable of deceiving the model can enhance the ranking position of the perturbed document. 

It is noteworthy that unlike term spamming techniques, these attack strategies can subtly manipulate document content in ways that are more imperceptible to both humans and machines, rendering them challenging to detect. 
For example, Chen et al. \cite{chen2023towards} propose a method that generates a pool of connection sentences by prompting a generative Language Model (LM) with a target pair of document and query. Then the most effective sentence~---~which promotes the ranking of the target document while maintaining coherence within the original document text~---~is selected and injected into the target document to increase its chance of exposure. In other work,
Wu et al. \cite{wu2023prada} introduced a word-level substitution method for attacking neural ranking models. Their proposed method pinpoints the key tokens within a document through the gradient of a surrogate model which are then substituted with their nearest neighbors, selectively enhancing the document's ranking if the substitution proves beneficial.

While these state-of-the-art methods have established a strong foundation and achieve significant attack success rate, many current methods encounter two key challenges, namely: (1) They depend heavily on surrogate models to generate adversarial documents, which requires a substantial amount of in-distribution training data obtained by querying the victim model. As a result, the attacking method often exhibit lack of robustness and a significant drop in attack performance metrics when models trained on easily accessible out-of-domain data are employed for crafting the adversarial documents. Moreover, this dependency not only reduces robustness across various victim models, but also limits their practicality in real-world scenarios where querying the victim model is not feasible or accessible. (2) They can generate adversarial documents that exhibit grammatical errors, nonsensical word sequences, and incoherent text fragments, rendering a considerable portion perceptible to both humans and machines that the document has been manipulated.

In response to these challenges, we present the Embedding Perturbation Rank Attack (\texttt{EMPRA}) method, which is a surrogate-agnostic method designed to execute adversarial black-box attacks on NRMs. \texttt{EMPRA} strategically manipulates \textit{sentence-level} embeddings to enhance the ranking of specific target documents. For sentence-level perturbations, \texttt{EMPRA} iteratively operates on the embedding representation of a document's sentences. This iterative process involves two key functions: (1) a \textit{transporter function}, which shifts sentence representations closer to the query context, and (2) a \textit{transformer function}, which converts the perturbed embedding representations into lexical form. The objective is to guide the sentences embeddings towards the context of the query while maintaining certain constraints that prevent substantial semantic deviation from the original sentences. After a set number of iterations, \texttt{EMPRA} generates adversarial text that not only encapsulate information from the original document's sentences but also exhibit semantic proximity to query-related information. Followed by this, the generated adversarial text is injected into the target document while preserving coherence and relevance, thus culminating in the production of a final adversarial document that is imperceptible to humans and machines. Unlike many baselines, \texttt{EMPRA} generates adversarial documents without relying on any specific surrogate NRM. This independence from surrogate model selection enhances its robustness, enabling it to adapt seamlessly to both In-Distribution (ID) and Out-of-Distribution (OOD) scenarios while ensuring appropriate attack performance against a diverse range of victim models.

To evaluate the efficacy of \texttt{EMPRA}, we conduct experiments utilizing the MS MARCO V1 passage collection \cite{nguyen2016ms} used by all prior works to attack NRMs. When targeting documents from the ranked list for diverse queries, \texttt{EMPRA} consistently outperforms state-of-the-art baselines, notably improving the ranking positions of target documents. Specifically, \texttt{EMPRA} outperforms the baselines by re-ranking almost 96\% of attacked documents that originally ranked 51-100 and 65\% of documents that ranked 996-1000 into the top 10. Furthermore, our experimental findings highlight the robustness of \texttt{EMPRA} across various victim NRMs, underscoring its performance reliability in real-world scenario attacks. Notably, \texttt{EMPRA} demonstrates an ability to  generate documents of high grammatical quality that remain imperceptible to human observations and machine. 

The key contributions of our work can be enumerated as follows:

\begin{itemize}

\item We propose a black-box adversarial attack method against neural ranking models that applies embedding perturbations on sentences within documents to generate adversarial documents that can outperform state-of-the-art baselines in terms of attack performance.

\item We report extensive attack experiments demonstrating that our method effectively ranks the adversarial documents in high positions. Since it is surrogate-agnostic and does not rely on surrogate models for adversarial document generation, our attack method is the most robust and effective against various victim models.

\item We demonstrate that \texttt{EMPRA} generates low-perplexity and fluent adversarial documents that can remain imperceptible under both human and automatic evaluations.
\end{itemize}

\section{Related Work}

In this section, we first provide an overview of adversarial attack research across various other domains and then introduce existing adversarial attack work specifically designed for text ranking models.

\subsection{Adversarial Attacks Across other Domains}
Since the emergence of Deep Neural Network (DNN) models, the research community has extensively studied their robustness to adversarial attacks across various fields such as computer vision \cite{akhtar2021advances,siddhant2019survey,akhtar2018threat,long2022survey}, recommender systems \cite{fan2022comprehensive,deldjoo2021survey,wang2024poisoning,wu2021triple,lin2020attacking}, and natural language processing \cite{goyal2023survey,qiu2022adversarial,BERTAttack,song2020adversarial,zou2019reinforced,ebrahimi2017hotflip,maimon2022universal}. 

Within the field of computer vision, adversarial attacks typically involve perturbing the representation of image inputs in imperceptible ways with the objective of misleading the prediction output of the victim model. These attacks have been explored in depth, and significant studies have highlighted the vulnerabilities of image classifiers to such adversarial attacks \cite{akhtar2021advances,siddhant2019survey,akhtar2018threat,long2022survey}. Various attacks such as the Fast Gradient Sign Method (FGSM) \cite{goodfellow2014explaining} have demonstrated how gradient-based perturbations can drastically manipulate a model's classification output and significantly reduce its accuracy. Building on FGSM, Carlini and Wagner \cite{carlini2017towards} introduced the C\&W attack, which employs an optimization framework that generates adversarial examples by minimizing perceptual perturbations while ensuring misclassification. This attacking strategy effectively exposes the limitations of defense mechanisms like defensive distillation \cite{papernot2016distillation} and often succeeds where simpler methods like FGSM fail by focusing on reducing the perceptual impact of the perturbation while still achieving the desired attack outcome. In addition, there are works that go beyond classification and focus on adversarial attacks on object detection systems and face recognition \cite{akhtar2021advances}. For instance, adversarial patch attacks introduce visible perturbations strategically placed within an image to mislead the object detection systems in identifying objects \cite{brown2017adversarial}. Similarly, adversarial attacks on face recognition systems aim to alter facial features imperceptibly in a way that misleads the model into making incorrect identifications \cite{sharif2016accessorize}.

In recommender systems, adversarial attacks exploit vulnerabilities by introducing fake user interaction data such as fake reviews, manipulated ratings, or fake user profiles to manipulate recommendation results of the system. These manipulations are often carried out through data poisoning attacks, where the goal of adversaries is to  inject malicious data to promote or demote specific items within the recommendation system output \cite{fan2022comprehensive,deldjoo2021survey}. For instance, \citet{chen2022knowledge} proposed a novel framework that leverages a knowledge graph to integrate auxiliary item attribute knowledge into a hierarchical policy network to enable the generation of high-quality fake user profiles designed to manipulate the recommendation outputs of the system effectively. \citet{fan2023adversarial} proposed a black-box attack framework that leverages transferable cross-domain user profiles from a source for crafting adversarial attack on the target domain through a locally trained surrogate model. 

In natural language processing, adversarial attacks can be categorized based on the granularity of the perturbations. These attacks include character-level, word-level, sentence-level, and multi-level approaches, with techniques ranging from simple character swaps and word substitutions to more complex paraphrasing or generative methods \cite{goyal2023survey,qiu2022adversarial}. For instance, HotFlip \cite{ebrahimi2017hotflip} introduced a white-box gradient-based character-level attack that identifies optimal perturbations in text by leveraging model gradients to craft minimal yet highly effective adversarial examples. TextFooler \cite{jin2020bert} employs word-level substitutions using semantically similar and grammatically correct replacements to mislead NLP models output while preserving the original input's fluency and coherence. BERT-Attack \cite{BERTAttack} is a contextualized adversarial attack method that leverages the masked language modeling capabilities of BERT \cite{devlin2018bert} to craft adversarial examples. It identifies and replaces vulnerable words with semantically consistent and grammatically correct substitutions and is capable of effectively misleading the victim model’s prediction outputs while maintaining input fluency and coherence. It is important to note that these attack methods are specifically designed for NLP tasks such as sentiment analysis and text classification, where the output consists of classification scores. In contrast, our work and related studies focus on adversarial attacks targeting text ranking models, which fundamentally rely on ranking relevance criteria between queries and documents, resulting in different outputs.

At a high level, adversarial attacks can be categorized into three types:

\begin{itemize}
\item White-box Attacks: The attacker has full access to the training data, model architecture, hyperparameters, and other details of the victim model \cite{ebrahimi2017hotflip,fang2020influence}. This type of attack is primarily used for research and development to build resilient models and is not practical in real-world scenarios, as attackers typically do not have extensive access to model settings. 

\item Gray-box Attacks: The attacker has partial access to the model's information, such as user reviews and ratings on platforms like Amazon \cite{wang2022gray,fang2018poisoning}. This makes gray-box attacks more practical and dangerous than white-box attacks.

\item Black-box Attacks: In many real-world contexts, the attack happens under a black-box scenario where the attacker has no information about the target model's settings or internal details. They can typically only query the target model to refine their attack strategies \cite{fan2023adversarial,song2020poisonrec}. This type of attack is the most realistic scenario for attackers targeting the neural network models used in commercial contexts.

\end{itemize}

\subsection{Adversarial Attacks on Text Ranking Models} 
The advancement of neural models in information retrieval has led to a paradigm shift from traditional term-frequency-based approaches, such as BM25 \cite{robertson1994some}, which rely on exact term matching between queries and documents, to neural models capable of capturing the semantic relationships between queries and documents. For instance, various studies have shown that fine-tuning transformer-based models such as BERT \cite{devlin2018bert} for information retrieval tasks significantly outperforms traditional methods \cite{nogueira2019passage,nogueira2020document,lin2022pretrained,pradeep2021expando}. This is primarily due to their ability to capture the semantic context of words within a query and document, along with their capacity to learn relevance by constructing an embedding space that brings the representations of queries and their relevant documents closer together while increasing the distance to irrelevant documents. Despite their impressive retrieval effectiveness, the core language models underlying these neural-based systems are vulnerable to adversarial attacks \cite{jin2020bert,BERTAttack}. These vulnerabilities can be exploited in different ways within the context of text ranking models, allowing attackers to manipulate the rankings by promoting or demoting specific documents in the list presented to users.

Recently, there has been growing attention towards assessing the robustness of the neural-based retrieval systems against adversarial attacks, particularly in the context of black-hat SEO and web spamming attacks \cite{patil2013search,gyongyi2005web}. Adversarial attacks in the search domain aim to manipulate a target document to deceive the model into ranking the perturbed document higher, thereby increasing its exposure to user search queries \cite{castillo2011adversarial}. These attacks pose significant threats to the fairness and integrity of search engines by exploiting vulnerabilities in ranking algorithms to undermine relevance criteria, distort fair ranking practices, and promote biased or misleading content.

Adversarial attack methods in search systems can be broadly categorized based on the type of target model: (1) attacks targeting retrieval models \cite{liu2023black,song2022trattack,zhong-etal-2023-poisoning} and (2) attacks targeting Neural Ranking Models (NRMs) \cite{wu2023prada,wang2022bert,liu2022order,chen2023towards}. Among these, adversarial attacks targeting NRMs are of particular interest and form the central focus of this work. Black-box adversarial attacks on NRMs can be further classified into three categories: (1) word-level-based attacks \cite{raval2020one,wu2023prada,wang2022bert}, (2) trigger-based generation attacks \cite{liu2022order,wang2022bert}, and (3) prompt-based attacks \cite{chen2023towards}.

\noindent\textbf{(1) Word-level-based} attacks target semantically important words in the document and replace them with semantically similar words that have closer representations to query within an embedding space. For instance, \citet{wu2023prada} use a surrogate model to detect important words within the target document and employ a greedy approach to replace those words with their nearest neighbors within the embedding perturbation space. \texttt{EMPRA} distinguishes itself by being a surrogate-agnostic attack method when generating adversarial documents and not being dependent on a specific model trained on either in-distribution or out-of-distribution data. As a result, it offers greater flexibility and adaptability in various attack settings.

\noindent \textbf{(2) Trigger-based} attacks aim to craft a short text and insert it into the document. Methods like Pairwise Anchor-based Trigger (PAT) \cite{liu2022order} add several trigger tokens at the beginning of the document using a ranking incentive objective equipped with fluency and semantic constraints to craft the adversarial document. In a similar study, Wang et al. \cite{wang2022bert} use HotFlip \cite{ebrahimi2017hotflip}, a gradient-based attacking technique, to add or replace tokens inside the target document to promote its ranking. Both \cite{liu2022order} and \cite{wang2022bert} use a surrogate model for generating adversaries that are injected into the target documents. \texttt{EMPRA} not only eliminates the need for a surrogate model but also generates adversarial texts that are semantically related to both the target query context and the original target document. This makes the final adversarial documents imperceptible to humans and machines by avoiding the use of irrelevant triggers and maintain the natural flow and coherence of the document. Besides, by incorporating semantically similar sentences related to both the target query context and the target original document, \texttt{EMPRA} demonstrates robustness against various victim models.

\noindent \textbf{(3) Prompt-based} attacking models prompt a generative LM to generate meaningful adversarial texts given a target document and a query that when are injected into the target document can boost its ranking for the target query. For instance, Chen et al. \cite{chen2023towards} propose an attack leveraging \texttt{BART} \cite{lewis2019bart}, a generative LM, which generates connecting sentences for the target document and query pair. By applying relevance and fluency constraints using a surrogate model and an LM, they inject the highest-scoring connection sentence into the document. 
\texttt{EMPRA} sets itself apart from \cite{chen2023towards} and others \cite{wu2023prada,liu2022order,wang2022bert} by not only considering the target query but also taking into account the top-ranked document when generating adversarial texts. This results in adversaries that are more effective in boosting the rankings of target documents, as they are semantically similar to the query, the context of the top-ranked document, and the target document itself, thereby enhancing their impact.

Various studies have emerged that leverage these categories of attack across various domains and contexts. For instance, Liu et al. \cite{liu2023topic} introduce a framework that  employs reinforcement learning with attacking actions drawn from \cite{wu2023prada} and \cite{liu2022order}, enabling the agent to perturb documents, elevating the target document's visibility for a set of semantically similar queries. A similar study \cite{liu2024multi} presents a framework utilizing reinforcement learning to orchestrate a diverse set of existing attacking methods, employing GPT-4 output fluency as the reward function at every state to craft adversarial documents. While these RL-based methods achieve slightly better attack performance compared to individual attack methods, they are significantly more time-consuming due to the complexity of RL and the computational demands of large language models. This may make them challenging to scale for practical real-world applications.

\section{Threat Model}

\subsection{Attack Objective}
Let $D^{\mathcal M_{V}}_{q}= [d_{1},d_{2},...,d_{m}]$ represent a list of $m$ ranked documents for a query $q$ from a collection of documents $\mathcal C$ by a {\em victim} neural ranking model $\mathcal M_{V}$, which is targeted by the attack. These documents are ordered according to the relevance scores assigned by the victim neural ranking model to each query-document pair, denoted as $f_{rel}(q, d_{j})$, where $j$ ranges from 1 to $m$, and it holds that $f_{rel}(q, d_{j}) < f_{rel}(q, d_{j-1})$. The attacker's objective is to design an adversarial threat model that applies perturbations $p$ to a target document $d$ within $D^{\mathcal M_{V}}_{q}$ to create an adversarial document ${d}^{\text{adv}}$.
The adversarial document ${d}^{\text{adv}}$ succeeds in the attack objective if the degree of perturbations $||p||$ applied to the target document results in a higher score with respect to the query, thus achieving a better (lower) ranking position, i.e.:
\begin{equation}
\text{Rank}(q, {d}^\text{adv}) < \text{Rank}(q, d),
\label{eq:rank}
\end{equation}
where $\text{Rank}(q, {d}^\text{adv})$ and $\text{Rank}(q, d)$ denote the positions of the adversarial document ${d}^\text{adv}$ and the target document $d$ in the ranking for the query $q$, respectively. Additionally, the core content similarity between the original document $d$ and the adversarial document ${d}^{\text{adv}}$ must meet a minimum threshold to prevent from semantic drift after perturbation and make sure ${d}^{\text{adv}}$ preserve the core content of the target document. The core content similarity threshold can be defined as:
\begin{equation}
CoreSim({d}, d^{\text{adv}}) \geqslant \lambda,
\label{eq:sim}
\end{equation}
where the core similarity function evaluates the key semantic elements to ensure that the core content intended to be exposed to the user remains aligned with the original content even after manipulation.

It is also important to make the perturbations in a manner that remains imperceptible to both humans and machines, while successfully deceiving the victim NRM. In particular, adding nonsensical or irrelevant phrases that degrade the readability of the adversarial document undermines the attack objective by decreasing user confidence and trust.

\subsection{Attacker’s Background Knowledge} \label{attackers_background_knowledge}
To align with real-world scenarios and consistent with prior studies on adversarial attacks against neural ranking models \cite{wu2023prada,liu2022order,chen2023towards}, our work focuses on the black-box adversarial attack setting. In this framework, the attack strategy employed by the attacker to craft the perturbed document ${d}^{\text{adv}}$ lacks access to any information regarding the victim neural ranking model, including its hyperparameters, gradients, and training data. As a result, the attacker can only query the victim neural ranking model $\mathcal M_{V}$ with a set of queries and use its output for constructing adversarial documents. 

To guide the attack process in a black-box setting, the attacker can adopt one of two strategies, categorized as \textit{surrogate-aware} or \textit{surrogate-agnostic} attack methods:

\begin{itemize}
\item \textbf{Surrogate-Aware Methods:}
Surrogate-aware methods rely on training a surrogate model, denoted as \(\mathcal{M}_S\), that mimics the behavior of the victim model. The surrogate model is trained using pseudo-relevance labels generated by querying the victim model and collecting its ranked output for a predefined set of queries. By learning to approximate the ranking criteria of the victim model, surrogate models enable the attacker to generate adversarial documents that exploit the victim model’s vulnerabilities. This approach is practical only when sufficient query access to the victim model is available, allowing for a close approximation of its behavior. However, surrogate-aware methods require substantial computational resources for training, and the process must be repeated if the victim model changes, making it costly and time-consuming. Additionally, adversarial documents crafted using a specific surrogate model may lack robustness when applied to other victim models, limiting their generalizability.

\item \textbf{Surrogate-Agnostic Methods:}
Surrogate-agnostic methods do not require explicit training on pseudo-relevance labels or extensive querying of the victim model. Instead, they leverage a generic neural ranking models, denoted as \(\mathcal{M}_G\), trained on publicly available datasets, to act as general relevance scoring models. Unlike surrogate-aware methods, surrogate-agnostic methods operate independently of the target victim model and offer a practical alternative in scenarios where access to the victim model is restricted or infeasible. This independence not only reduces the computational overhead but also enhances the adaptability and robustness of the attack. By using models that are not tied to any specific victim model, surrogate-agnostic methods ensure broader applicability, making them well-suited for diverse attack scenarios without requiring retraining or adjustments when the victim model changes.
\end{itemize}

\begin{algorithm}[t!]
\small
\caption{EMPRA}
\label{alg:empra}
\KwIn{
    Query $q$; 
    Target Document $d = [S_{1}, \dots, S_{|d|}]$; 
    Anchor Texts $\mathcal{A} = [A_1, \dots, A_{k}]$; 
    Embedding Function $\embed{\cdot}$; 
    Transporter Function $\mathcal{T}(\cdot,\cdot)$; 
    Transformer Function $\Pi(\cdot)$; 
    Generic NRM $\mathcal{M}_{G}$;
}
\KwParam{
    Step size $\eta$; 
    Perturbation bound $\epsilon$; 
    Max Iterations $N$; 
    Interpolation Coefficient $\alpha$; 
}
\KwOut{
    Adversarial Document $d^{\mathrm{adv}}$
}

\BlankLine
\textbf{Stage 1: Adversarial Text Generation}

\textbf{Initialize} ${T}_{\text{adv}} \leftarrow$ []

\ForEach{$S \in d$}{
    \ForEach{$A \in \mathcal{A}$}{
        \textbf{Initialize} $T^{(0)} \leftarrow S$ \\
        \textbf{Initialize} $\embed{S}^{(0)} \leftarrow \embed{S}$ \\
        
        \For{$t = 0 \dots N-1$}{
        $\embed{S}^{(t+1)} = \mathcal{T}(\embed{S}^{(t)}, \embed{A})$ \\
            
            $T^{(t+1)} \leftarrow \Pi\bigl(\embed{S}^{(t+1)}\bigr)$
        }
        
        ${T}_{\text{adv}} \leftarrow {T}_{\text{adv}} \cup \{\,T^{(N)}\}$
    }
}
\BlankLine
\textbf{Stage 2: Adversarial Document Construction}

\textbf{Initialize} $d^{\mathrm{adv}}_{\mathrm{best}} \leftarrow d$, 
$\mathit{bestScore} \leftarrow -\infty$.

\ForEach{$T_i \in {T}_{\text{adv}}$}{
    \For{$p = 0 \dots |d|$}{
        \eIf{$p = 0$}{
           $\displaystyle d^{\mathrm{adv}}_{i,p} \leftarrow T_i \oplus d$ 
        }{
            \eIf{$p = |d|$}{
                $\displaystyle d^{\mathrm{adv}}_{i,p} \leftarrow d \oplus T_i$
            }{
                
                $\displaystyle d^{\mathrm{adv}}_{i,p} 
                    \leftarrow d_{1}^{p} \oplus T_i \oplus d_{p+1}^{|d|}$
            }
        }

        $\displaystyle C_{\text{coh}} \leftarrow \text{Evaluate via } f_{\text{nsp}}(q, d^{\mathrm{adv}}_{i,p})$

        $\displaystyle C_{\text{rel}} \leftarrow \text{Calculate relevance score using $\mathcal{M}_{G}$ via } f_{\text{rel}}(q, d^{\mathrm{adv}}_{i,p})$

        $\displaystyle Score_{\text{interp}} \leftarrow \alpha \cdot C_{\text{coh}} + (1 - \alpha) \cdot C_{\text{rel}}$
        
        \If{$Score_{\text{interp}} > \mathit{bestScore}$}{
            $\displaystyle \mathit{bestScore} \leftarrow Score_{\text{interp}}$ \\
            $\displaystyle d^{\mathrm{adv}}_{\mathrm{best}} \leftarrow d^{\mathrm{adv}}_{i,p}$
        }
    }
}
\BlankLine

\textbf{Output:} 
$\displaystyle d^{\mathrm{adv}} \leftarrow d^{\mathrm{adv}}_{\mathrm{best}}$.

\end{algorithm}

\section{Proposed Method}\label{sec:proposed_method}
In this section, we introduce our proposed black-box adversarial attack method, referred to as \texttt{EMPRA}, a surrogate-agnostic attack strategy designed to manipulate the content of a target document with the goal of deceiving the victim neural ranking model into ranking the adversarial document higher in the rank list. \texttt{EMPRA} consists of two key stages: (1) \textit{Adversarial Text Generation}; and, (2) \textit{Adversarial Document Construction}. These stages integrate sentence-level embedding perturbations with strategic document construction to achieve an effective and imperceptible adversarial attack.

The first stage focuses on adversarial text generation, where the goal is to generate adversaries through a transition from discrete lexical modifications to more continuous adjustments in the embedding space. \texttt{EMPRA} leverages a transporter function to iteratively perturb the embedding of each sentence in the target document, moving it closer to an anchor text that is highly relevant to the query. The transformer function then maps the perturbed embedding back into lexical form. This iterative process generates a set of adversarial texts that are capable of boosting the ranking position of the target document and successfully manipulate the victim model.

The second stage involves constructing the adversarial document by strategically integrating the generated adversarial texts into the original document. Each adversarial text is evaluated for semantic coherence, measured using a Next Sentence Prediction (NSP) function, and query-document relevance, assessed through a generic neural ranking model. The final adversarial document is selected based on an interpolated scoring mechanism that balances coherence and relevance, ensuring that the document is both highly effective at deceiving the victim model and indistinguishable from natural text.

Our proposed adversarial attack method, \texttt{EMPRA}, is detailed in Algorithm~\ref{alg:empra}, which presents the pseudo-code outlining the steps for generating adversarial documents in the black-box attack setting.

\subsection{Adversarial Text Generation}
The adversarial text generation process is the core of our proposed attack method, with the goal of finding a set of text segments that, when appended to the target document, can boost its ranking in the rank list. Let $d$ represent a target document from $D^{\mathcal M_{V}}_q$, which our proposed attacking model $\aleph$ would like to manipulate so that it achieves a higher ranking for query $q$. Let $d^*$ denote the document currently ranked highest for query $q$. Also, let $\mathcal S_{d} = [S_{1},S_{2},...,S_{|d|}]$ represent $d$ as a sequence of sentences where $|d|$ is the total number of sentences in $d$. Also, let $\mathcal A = [A_1, A_2, ..., A_k]$ represent a collection of {\em anchor texts} that provide pertinent context related to 
$q$. These anchor texts can be defined as: 1) the query itself, 2) the top-ranked document, or 3) the most similar sentence from the top-ranked document to the target corresponding sentence in $d$. In order to generate a set of adversarial text that boost the ranking of $d$, our proposed attacking model $\aleph$ considers $\mathcal S_{d}$ and $\mathcal A$ to generate adversarial texts for $d$ as follows:
\begin{equation}
{T}_{\text{adv}} = \aleph(\mathcal S_{d}, \mathcal A),
\end{equation}
where ${T}_{\text{adv}} = [T_1, T_2, ..., T_m]$ consists of a set of adversarial text generated by $\aleph$. 
We note that the changes in the lexical form are limited and discrete, which increases the likelihood of deviating significantly from the original content. In contrast, working in the embedding space allows for more continuous adjustments, providing greater flexibility for slight perturbations. This enables a better balance between maintaining relevance and preserving semantic meaning. Therefore, instead of operating in the lexical space, we transition to the embedding space to achieve this balance. As such, in order to generate adversarial texts, the attacking model $\aleph$ leverages two components that work together in tandem, namely (\textbf{1}) a \textit{transporter function} $\mathcal{T(.)}$; and, (\textbf{2}) a \textit{transformer function} $\Pi(.)$.

Let $\embed{.}$ be the embedding function that maps target document sentences $\mathcal S_{d}$ and anchor texts $\mathcal A$ to their corresponding fixed-length vector representation within a high-dimensional embedding space. The goal of the transporter function $\mathcal T$ is to manipulate the target sentence embedding representation $\embed{S}$ within the embedding space to align it more closely with a target anchor vector representation $\embed{A}$.
The transformer function $\Pi$ is then responsible to transform the perturbed embedding representation to lexical form. The adversarial attack process involves iteratively adjusting the embedding representation of the document sentences to converge towards the desired target anchor texts, thereby enhancing the similarity score between the sentences and anchor texts. 

Given the sentence embedding $\embed{S}$ and the anchor embedding $\embed{A}$, the transporter function calculates the new coordinates of the sentence embedding representation through $\embed{S}^{(t+1)} = \mathcal{T}(\embed{S}^{(t)}, \embed{A})$, where $t$ represents the iteration step in the adversarial text generation process. The transformer function then maps the embedding representation to its corresponding lexical form, defined as $T^{(t+1)} = \Pi(\embed{S}^{(t+1)})$. Each iteration of our approach can be defined as the following two steps:

\begin{equation}
\begin{aligned}
T^{(t+1)} = \Pi\Bigg(&\embed{S}^{(t)} + \eta \cdot \operatorname{clip}\left(
    \frac{\partial}{\partial \mathbf{S}} 
    \left( \frac{\embed{S}^{(t)} \cdot \embed{A}}{\left\|\embed{S}^{(t)}\right\|_2 \left\|\embed{A}\right\|_2} \right), 
    -\epsilon, \epsilon
\right)\Bigg), \\
&t = \{0, 1, 2, \ldots, N-1\}.
\end{aligned}
\end{equation}

Here, $\eta$ denotes the step size of embedding perturbation and $\operatorname{clip}$ ensures the perturbation is within the specified $\epsilon$ bounds i.e., to retain the information of the original sentence. 
Such constraints are imposed on the search space to ensure that the representation of the perturbed embedding $\embed{S}$ does not deviate significantly from its original embedding state $\embed{S}^{(0)}$. This is achieved by limiting the magnitude of perturbations to ensure they fall within an $L_{\infty}$ distance of the original representation with a specified radius of $\epsilon$. After $N$ iterations, the adversarial textual representation of the original sentence $S$ that is closer to the anchor text but still close to the original representation of the sentence is obtained.

The transformer function $\Pi$ iteratively enhances a textual hypothesis $H^{(i)}$, which is an intermediate version of a sentence generated during the attack process. At each iteration, the goal is to diminish the divergence between its associated embedding $\embed{H^{(i)}}$ and a desired embedding target $\embed{S}^{(t+1)}$ during the transformation. This process gradually refines the lexical representation, aiming to minimize the distance between $\embed{H^{(i)}}$ and $\embed{S}^{(t+1)}$. This approach allows for finding progressively closer approximations to the target embedding, utilizing the transformer function to generate initial hypotheses and iteratively refining it. 
Having the transporter and transformer functions working in tandem, for every sentence $S \in \mathcal S_{d}$ and every anchor $A \in \mathcal A$, the attack model can generate adversaries as follows:

\begin{equation}
T^{(t+1)} = \Pi\big(\mathcal{T}(\embed{S}^{(t)}, \embed{A})\big), 
\quad t = \{ 0, 1, 2, \ldots, N-1 \}.
\end{equation}

Where  $T^{(0)} = S$. This iterative process generates a set of adversaries ${T}_{\text{adv}}$ for document $d$ without relying on any specific surrogate model, making the adversarial text generation independent of the surrogate model selection.

\subsection{Adversarial Document Construction}
To construct the final adversarial document ${d}^{\text{adv}}$, our proposed attacking model $\aleph$ injects each of the generated adversarial text $T$ into different positions within the target document $d$ to construct the adversarial document ${d}^{\text{adv}}$. This approach identifies the most effective adversarial text in enhancing both the relevance score and fluency of the target document. This balance is crucial for maintaining the fidelity of the perturbed document and achieving an optimal trade-off between its effectiveness at tricking the victim neural ranking model $\mathcal M_{V}$ and imperceptibility to human judges and automated systems. To this end, the insertion operation $\mathcal{I}(d, T, p)$ places each $T$ from ${T}_{\text{adv}}$ at position $p$ of the target document to build different candidates for ${d}^{\text{adv}}$ as follows:

\begin{equation}
d^{\text{adv}}_{i, p} = 
\begin{cases} 
T_i \oplus d & \text{if } p = 0 \\ 
d_1^p \oplus T_i \oplus d_{p+1}^{|d|} & \text{if } 0 < p < |d| \\ 
d \oplus T_i & \text{if } p = |d|
\end{cases}
\end{equation}

where $\oplus$ denotes concatenation, and $d_a^b$ represents the sub-sequence of sentences from $S_a$ to $S_b$.  Given different variations of adversarial candidates, the effectiveness and coherence of each candidate $d^{\text{adv}}_{i, p}$ are quantified by two principal metrics, namely 1) \textit{semantic coherence}; and, 2) \textit{relevance to the query}. In order to evaluate semantic coherency, a coherence score $C_{\text{coh}}$ is calculate using pre-trained BERT Next Sentence Prediction (NSP) function denoted as $f_{nsp}$. This function assesses the compatibility between adjacent document sentences. For an adversarial sentence $T_i$ inserted at position $p$, the coherence score is defined as follows:

\begin{equation}
C_{\text{coh}}(d^{\text{adv}}_{i, p}) = 
\begin{cases} 
f_{\text{nsp}}(T_i, d) & \text{if } p = 0, \\ 
f_{\text{nsp}}(d, T_i) & \text{if } p = |d|, \\ 
\frac{1}{2} \Big[ f_{\text{nsp}}(d_{1}^{p}, T_i \oplus d_{p+1}^{|d|}) + 
f_{\text{nsp}}(d_{1}^{p} \oplus T_i, d_{p+1}^{|d|}) \Big] & \text{if } 0 < p < |d|.
\end{cases}
\label{eq:coherence}
\end{equation}

To calculate the relevance score between each adversarial candidate and the query in the black-box setting, our proposed surrogate-agnostic attacking model utilizes the relevance scoring function of a generic neural ranking model $M_{G}$, as follows:
\begin{equation}
C_{\text{rel}}(q, d^{\text{adv}}_{i, p}) = f_{\text{rel}}(q, d^{\text{adv}}_{i, p}).
\label{eq:relevance_equation}
\end{equation}

The generic model can be any neural ranking model other than the victim model itself, trained on in-distribution or out-of-distribution data to learn the relevance criteria. It is worth mentioning that this makes the process of constructing the adversarial document ranker-agnostic, as it does not necessarily depend on a specific surrogate model. This design choice ensures greater generalizability and flexibility of the proposed attack method across different ranking systems.

In order to select the best adversarial candidate, an interpolated score $Score_{\text{interp}}$ is computed to balance the trade-off between semantic coherence and query-document relevance as shown below:
\begin{equation}
Score_{\text{interp}}(q, d^{\text{adv}}_{i, p}) = \alpha \cdot C_{\text{coh}}(d^{\text{adv}}_{i, p}) + (1 - \alpha) \cdot C_{\text{rel}}(q, d^{\text{adv}}_{i, p}),
\label{eq:interpolation}
\end{equation}
where $\alpha$ is the interpolation coefficient, and both $C_{\text{coh}}$ and $C_{\text{rel}}$ are normalized to be within the range of [0, 1]. The adversarial document $d^{\text{adv}}$ for the target document $d$ would be the candidate with the highest $Score_{\text{interp}}$, ensuring a balanced approach that maximizes attack efficacy while maintaining semantic coherence, thereby reducing the risk of detection. 

\section{Experimental Setup}
We make our code and experimental data publicly available at {\url{https://github.com/aminbigdeli/EMPRA}}.

\subsection{Datasets}
\subsubsection{Benchmark Datasets.}
Similar to previous studies \cite{chen2023towards,wu2023prada,liu2022order,wang2022bert}, we utilize the MS MARCO V1 Passage Collection \cite{nguyen2016ms}, which encompasses 8.8 million passages. This collection includes over 500,000 training queries, a small validation set (dev small) with 6,980 queries, and a small eval set with 6,837 queries. An adequate number of training, validation, and test queries make this dataset suitable for training both victim and surrogate NRMs, as well as for evaluating the performance of attack methods. Additionally, A processed version of the Natural Questions (NQ) dataset \cite{kwiatkowski2019natural} as prepared by \cite{karpukhin2020dense}, the SQuAD2.0 dataset \cite{rajpurkar-etal-2018-know}, the Stanford Natural Language Inference (SNLI) dataset \cite{bowman-etal-2015-large}, and the Community QA (CommonQA) dataset \cite{ni-etal-2022-sentence} were used to serve as out-of-distribution training datasets for training generic NRMs. This allowed us to explore the stability and robustness of \texttt{EMPRA} and compare it against various attacking methods. 

\subsubsection{Target Queries and Documents.}
To evaluate the performance of the attack strategies, we follow the approach of~\citet{chen2023towards}. First, we randomly selected 1,000 queries from the MS MARCO dev set. For each query, 10 documents were targeted from the re-ranked list generated by the target victim model from the top-1000 documents initially retrieved by BM25. These documents were classified into `Easy-5' and `Hard-5' groups based on the anticipated difficulty of boosting their rankings into the top-10 or top-50 ranking positions. 
The Easy-5 documents were randomly chosen from each 10-document segment within positions 51-100. The Hard-5 documents comprised of the last five ranked documents by the victim model, occupying positions 996 to 1,000. As a result, each attacking method was tasked with producing adversarial documents for a total of 10,000 documents. Additionally, in line with prior work~\cite{chen2023towards,liu2022order}, we include a set of `\textit{Mixture}' target documents for comprehensive analysis. This set comprises 32 target documents sampled from both Easy and Hard categories to ensure a balanced representation of varying difficulty levels. The Mixture target documents are specifically utilized for human-based evaluations and metrics requiring high computational costs, enabling an efficient and comprehensive analysis of the attack methods.

\subsection{Evaluation Metrics}

\subsubsection{Attack Performance}
Similar to previous studies \cite{chen2023towards,wu2023prada,liu2022order}, we consider a set of comprehensive attack performance metrics each capturing unique aspects of the effectiveness of our proposed adversarial method and baselines on document rankings.

\noindent \textbf{Attack Success Rate (ASR).}
This metric evaluates the effectiveness of an attack by calculating the proportion of targeted documents that achieve a higher ranking post-attack compared to their original position, averaged across all queries:

\[
ASR = \frac{1}{|Q|} \sum_{q \in Q} \frac{1}{N_q} \sum_{i=1}^{N_q} \mathds{1}\left( Rank_{\mathcal{R}_{q}}(d^{\text{adv}}_{i}) < Rank_{\mathcal{R}_{q}}(d_{i}) \right).
\]

Here, \(Q\) is the set of evaluated queries, \(N_q\) represents the number of targeted documents for a specific query \(q\), and \(\mathds{1}(\cdot)\) is the indicator function, which returns 1 if the condition inside is true, and 0 otherwise. A higher attack success rate, as measured by the proportion of adversarial documents achieving improved rankings, indicates a more effective attack strategy.

\noindent \textbf{Boosted top-k.}
Boosted top-k, represented as $\%r\leq k$, calculates the proportion of adversarial documents \(d^{\text{adv}}\), originally ranked outside the top-\(k\), that are moved into the top-\(k\) positions post-attack, averaged over all queries:

\[
\%r \leq k = \frac{1}{|Q|} \sum_{q \in Q} \frac{1}{N_q} \sum_{i=1}^{N_q} \mathds{1}\left( Rank_{\mathcal{R}_{q}}(d_{i}) > k \land Rank_{\mathcal{R}_{q}}(d^{\text{adv}}_{i}) \leq k \right).
\]

This metric highlights the attack's capacity to significantly alter the visibility of lower-ranked documents. We report the performance of various attack methods using $\%r\leq 10$ and $\%r\leq 50$ to measure the percentage of adversarial documents boosted into the top-10 and top-50 positions, respectively.

\noindent \textbf{Average boosted ranks (Boost).}
This metric measures the average improvement in rankings for adversarial documents \(d^{\text{adv}}\), averaged across all queries:

\[
Boost = \frac{1}{|Q|} \sum_{q \in Q} \frac{1}{N_q} \sum_{i=1}^{N_q} \left( Rank_{\mathcal{R}_{q}}(d_{i}) - Rank_{\mathcal{R}_{q}}(d^{\text{adv}}_{i}) \right).
\]

It reflects how effectively the attack method can elevate the ranking positions of target documents and highlights its impact on improving their visibility within the search results.

\subsubsection{Quality and Naturalness}
To fully evaluate the quality and naturalness of the adversarial documents generated by our attacking method and baselines, we employed six different metrics, following the approach of previous work \cite{chen2023towards,wu2023prada,liu2022order}. Among these metrics, Perplexity and Readability were measured at scale across all generated documents for all attacking strategies. Due to computational costs and the limited availability of human annotators,the remaining metrics were evaluated exclusively on the `Mixture' target documents.

\noindent \textbf{Perplexity.} To evaluate the fluency of the generated adversarial documents, we utilize a pre-trained \texttt{GPT-2} model \cite{radford2019language} to compute the language model perplexity. Perplexity is defined as:

\[
\text{Perplexity}(d^{\text{adv}}) = \exp\left(-\frac{1}{N} \sum_{i=1}^{N} \log P(w_i | w_1, w_2, \dots, w_{i-1})\right),
\]

where \(N\) is the total number of words in the adversarial document \(d^{\text{adv}}\). A lower perplexity value indicates higher fluency, as it implies that the language model assigns higher probabilities to the sequence of words in the adversarial document. This metric is critical for assessing whether the adversarial text maintains linguistic coherence and naturalness and minimizes its likelihood of being detected as manipulated content.

\noindent \textbf{Readability.} To evaluate the readability of generated adversarial documents, we use the Dale-Chall readability score \cite{dale1948formula}, which estimates text complexity based on familiar word usage and sentence structure. This metric compares the text against a list of 3,000 commonly known words and uses the proportion of unfamiliar words and average sentence length to determine the grade level needed for comprehension. Higher scores indicate greater difficulty and help assess whether adversarial modifications impact the text’s accessibility and readability.

\noindent \textbf{Grammar Assessment}
We employ Grammarly \cite{grammarly2023} to assess the quality of the adversarial documents by submitting them to Grammarly website to obtain their overall quality score. This score reflects the grammatical accuracy, sentence structure, and stylistic appropriateness of the text. Therefore, we can quantify the impact of adversarial perturbations on the grammatical quality of the generated documents and determine whether the perturbations introduce noticeable errors.

\noindent \textbf{Linguistic Acceptibility}
For assessing linguistic acceptability, we utilize a language classification model \cite{warstadt2019neural} trained to evaluate whether a text adheres to standard linguistic norms. This model determines if the adversarial document maintains proper syntax, semantics, and overall coherence to meet the linguistic standards of natural language.

\noindent \textbf{Human Evaluation.} This aspect assesses the imperceptibility and fluency of adversarial documents generated by different methods. Annotators are presented with both original and adversarial documents and are asked to perform two key tasks: (1) identify whether the documents appear normal and imperceptible, meaning they do not exhibit obvious signs of adversarial manipulation; and (2) rate the fluency of the documents, reflecting their coherence, grammatical correctness, and naturalness. This evaluation ensures that adversarial documents meet real-world standards of imperceptibility and fluency while preserving user confidence and trust.

\subsection{Models}
\label{NRM_models}

\subsubsection{Victim NRMs}
Consistent with prior research \cite{liu2022order,chen2023towards}, we select the \texttt{msmarco-MiniLM-L-12-v2}\footnote{\url{https://huggingface.co/cross-encoder/ms-marco-MiniLM-L-12-v2}} model as our primary victim black-box neural ranking model, denoted as $\mathcal{M}_V$. This cross-encoder-based ranker is fine-tuned on the MS MARCO training set, leveraging \texttt{MiniLM} \cite{wang2020minilm} as its foundational language model for learning query-document semantic mapping. The model has demonstrated high retrieval effectiveness in terms of Mean Reciprocal Rank (MRR@10), as evidenced in Table \ref{table:mrr}. 

To assess the robustness of different attacking strategies across various victim NRMs, we extend our investigation to include two additional models: \texttt{ms-marco-electrabase}\footnote{\url{https://huggingface.co/cross-encoder/ms-marco-electra-base}} \cite{reimers2019sentence}, and \texttt{DistilRoBERTa-base} \cite{sanh2019distilbert}. Both models are fine-tuned on the MS MARCO training dataset and are based on distinct underlying language models compared to the primary victim NRM. Notably, the surrogate NRMs are not trained with pseudo-relevance feedback from these models. This setup enables a comprehensive evaluation of the effectiveness and robustness of different attack methods when targeting these additional victim NRMs.

\subsubsection{Surrogate NRMs}
Surrogate NRMs are employed as foundational components for executing surrogate-aware attack scenarios by interacting with the victim model through query submissions, leveraging the attacker's knowledge as outlined in Section \ref{attackers_background_knowledge}, to approximate the victim model's behavior effectively. Consequently, varying numbers of queries from MS MARCO eval small dataset are employed to query the primary victim model $\mathcal{M}_V$ and train distinct surrogate models based on the re-ranked list of documents generated by the victim model. As proposed by the authors in \cite{chen2023towards}, two query sets are utilized for this purpose, leading to the training of two surrogate models based on the pre-trained \texttt{BERT-base} \cite{devlin2018bert} using different query quantities:

\begin{itemize}
\item $\mathcal{M}_{S_1}$: Trained on the full set of 6,837 eval small queries from MS MARCO, serving as an ID surrogate model. 

\item $\mathcal{M}_{S_2}$: Trained on a smaller subset of 200 eval small queries from MS MARCO, also serving as an ID surrogate model with considerably less number of training queries.

\end{itemize}

Table \ref{table:mrr} compares the retrieval effectiveness of these two surrogate models against the primary target victim model $\mathcal{M}_V$ on the MS MARCO dev small set. As shown in the table, $\mathcal{M}_{S_1}$ is identified as the best-performing surrogate model, denoted as $\mathcal{M}_{S_{\text{best}}}$, due to its highest imitation capability, performing closely to the victim model. In contrast, $\mathcal{M}_{S_2}$ provides a less effective alternative which reflects the impact of its smaller training data. Both models serve as effective ID surrogates, facilitating the adversarial attack process by offering relevance predictions within the same data distribution as the victim model.

\begin{table}[!t]
\centering

\caption{The retrieval effectiveness (MRR@10) of the first-stage retriever (BM25), the primary victim model ($\mathcal M_V$), the surrogate models ($\mathcal M_{S_1}$ and $\mathcal M_{S_2}$), and the generic models ($\mathcal M_{G_{1-4}}$) on the dev small set of MS MARCO.}

\begin{tabular}{lccccccccc}
\hline
\\[-0.9em]
Model & BM25 & $\mathcal M_V$ & $\mathcal M_{S_1}$ & $\mathcal M_{S_2}$ & $\mathcal M_{G_1}$& $\mathcal M_{G_2}$& $\mathcal M_{G_3}$ & $\mathcal M_{G_4}$ \\ [0.2em] \hline  \\[-0.9em]
MRR@10 & 18.4 & 39.5 & 37.0 & 23.0 & 19.6 & 16.6 & 18.9 & 21.0 \\ [0.2em] \hline
\end{tabular}
\label{table:mrr}

\end{table}

\subsubsection{Generic NRMs}

Generic NRMs serve as key components in implementing surrogate-agnostic attack scenarios, as detailed in Section \ref{attackers_background_knowledge}. Unlike surrogate models, generic NRMs do not depend on pseudo-relevance labels generated through querying the victim model for training a neural ranking model. Instead, these models are pre-trained models that capture the concept of relevance by fine-tuning on diverse datasets, learning general principles of query-document matching. Generic models serve as an alternative approach in the black-box attack setting, eliminating the need for extensive querying of the victim model and enabling attack strategies that are independent of the victim model's specific behavior.

We utilize a variety of pre-trained embedding models that have been fine-tuned on publicly available datasets commonly used for training neural ranking models. These models are utilized as generic NRMs, designed to represent out-of-distributions scenarios to ensure the robustness and adaptability of \texttt{EMPRA}. The selected models are as follows:

\begin{itemize}
\item $\mathcal{M}_{G_1}$: A \texttt{mpnet-base} model \cite{song2020mpnet} fine-tuned on a set of questions and answers from the NQ dataset\footnote{\url{https://huggingface.co/tomaarsen/mpnet-base-nq}}.

\item $\mathcal{M}_{G_2}$: A \texttt{distilbert-base} model \cite{sanh2019distilbert} fine-tuned on the SQuAD2.0 dataset\footnote{\url{https://huggingface.co/Pennywise881/distilbert-base-uncased-mnr-squadv2}}.

\item $\mathcal{M}_{G_3}$: A \texttt{T5} model fine-tuned on the CommonQA and NLI datasets to create a sentence embedding model, referred to as \texttt{sentence-t5-base} \cite{ni-etal-2022-sentence}\footnote{\url{https://huggingface.co/sentence-transformers/sentence-t5-base}}.

\item $\mathcal{M}_{G_4}$: A \texttt{BERT-base} model fine-tuned on the NQ dataset, following the methodology outlined in \cite{chen2023towards}.

\end{itemize}

The retrieval effectiveness of the generic models on dev small is also reported in Table \ref{table:mrr}. Among these models, $\mathcal{M}_{G_4}$ is identified as the best-performing generic model, denoted as $\mathcal{M}_{G_{\text{best}}}$, in terms of retrieval effectiveness, while $\mathcal{M}_{G_2}$ exhibits the lowest performance. By employing these generic models as out-of-distribution models within the \texttt{EMPRA} framework, we can thoroughly assess the robustness and adaptability of our attack method.

\begin{figure}[t]
\centering
\begin{tcolorbox}[width=0.99\linewidth,colback=gray!5!white,colframe=black!75!black]
\textbf{Instruction:}
You are an expert assistant in the field of information retrieval. Given the query and the document below, perform an adversarial attack on the document to make it rank high for the query,
while maintaining the original content and structure of the document. Simply generate the perturbed document without explanation.
  \newline
  \newline
  Query: \{query\}
  \newline
  Document: \{document\} 
  \newline
  Perturbed Document:
\end{tcolorbox}
\caption{Prompt used by \texttt{GPT-4} for generating adversarially perturbed documents.}
\label{fig:llm_prompt}
\end{figure}

\subsubsection{Baselines}
To demonstrate the performance of our attacking method, we conduct a comparative study against state-of-the-art baseline methods across word-level-based, trigger-based, and prompt-based categories, along with an LLM-based baseline. The following methods serve as our benchmarks: \texttt{Query+} \cite{liu2022order} is a simple baseline that adds the query at the beginning of the target document. While the query could be inserted at any position within the document, placing it at the beginning yielded the most effective attack results. \texttt{GPT-4} (\texttt{gpt-4-1106-preview}) is employed to generate an adversarial document given the target query and original document. To achieve this, we design a prompt that instructs \texttt{GPT-4} to create an adversarial version of the document, while preserving its original content and structure, as shown in Figure \ref{fig:llm_prompt}. \texttt{PRADA} \cite{wu2023prada} detects important terms in the target document using the surrogate model and replaces at-most 20 tokens with their synonyms within an embedding space. \texttt{PAT} \cite{liu2022order} adds trigger words with the max length of 12 at the beginning of the target document. This method leverages a surrogate model in an incentivized manner to investigate whether the addition of these words enhances the document's ranking. \texttt{Brittle-BERT} \cite{wang2022bert} also adds trigger words at the beginning of the target document with the max length of 12. \texttt{IDEM} \cite{chen2023towards} generates 500 connection sentences using \texttt{BART} \cite{lewis2019bart} with the max length of 12 and selects the best one in terms of the relevance and fluency trade-off in order to inject it into the original document for creating the adversarial document. The authors reported that adding sentences longer than 12 will maintain the attack performance at almost the same level.

\subsubsection{Implementation Details} 
For the implementation of the transporter function, 
we employ an $L_{\infty}$ distance with a radius ($\epsilon$) of 0.01, a step size of 0.1, and 25 iterations to guide the movement of sentence embeddings towards the anchors. Furthermore, the interpolation coefficient $\alpha$ in Equation \ref{eq:interpolation} is set to 0.5. The impact of the number of iterations and $\alpha$ on the attack performance is investigated in the next section.

\section{Results and Findings}
In order to demonstrate the effectiveness of our proposed \texttt{EMPRA} method, we report and analyze the experimental results through the lens of the following research questions:

\begin{itemize}
    \item \textbf{RQ1:} What is the impact of using various surrogate/generic NRMs on the attack performance of \texttt{EMPRA}?
    \item \textbf{RQ2:} How does the attack performance of \texttt{EMPRA} compare to baseline methods in both in-distribution (ID) and out-of-distribution (OOD) settings?
    \item \textbf{RQ3:} How does the attack performance of \texttt{EMPRA} compare to baseline methods across various victim models?
    \item \textbf{RQ4:} How do different hyper-parameter settings affect the attack performance and semantic coherence of adversarial documents generated by \texttt{EMPRA}?
    \item \textbf{RQ5:} How well do the adversarial documents generated by \texttt{EMPRA} maintain quality, naturalness, and linguistic acceptability compared to baseline methods, and to what extent can they evade detection by human evaluators and automated detection systems?
\end{itemize}

\begin{table*}[t]
\centering

\caption{Attack performance of \texttt{EMPRA} across six different neural ranking models used for generating adversarial documents targeting the primary victim model $\mathcal{M}_V$.}

\scalebox{0.96}{
\begin{tabular}{ccccccccccccc}
\hline \\[-1em]

& & \multicolumn{5}{c}{Easy-5} & & \multicolumn{5}{c}{Hard-5} \\ [0.2em] \cline{3-7} \cline{9-13}  \\[-0.9em]
Model & Fine-tune data & ASR & $\%r \leq 10$ & $\%r \leq 50$ & Boost & PPL$\downarrow$ & & ASR & $\%r \leq 10$  & $\%r \leq 50$ & Boost & PPL$\downarrow$ \\ [0.2em]  \hline \\[-0.9em]
\multicolumn{1}{c}{$\mathcal M_{S_1}$} & MS MARCO & \multicolumn{1}{c}{99.9} & \multicolumn{1}{c}{95.6} & \multicolumn{1}{c}{99.8} & \multicolumn{1}{c}{72.5} & \multicolumn{1}{c}{34.4}  & & \multicolumn{1}{c}{99.9} & \multicolumn{1}{c}{64.9} & \multicolumn{1}{c}{87.0} & \multicolumn{1}{c}{948.4} & \multicolumn{1}{c}{47.1} \\ [0.2em]
\multicolumn{1}{c}{$\mathcal M_{S_2}$} & MS MARCO &\multicolumn{1}{c}{99.4} & \multicolumn{1}{c}{83.1} & \multicolumn{1}{c}{98.5} & \multicolumn{1}{c}{68.6} & \multicolumn{1}{c}{35.4} && \multicolumn{1}{c}{99.5} & \multicolumn{1}{c}{47.5} & \multicolumn{1}{c}{73.2} & \multicolumn{1}{c}{909.9} & \multicolumn{1}{c}{50.3} \\ [0.2em]
\multicolumn{1}{c}{$\mathcal M_{G_1}$} & NQ &\multicolumn{1}{c}{99.4} & \multicolumn{1}{c}{85.2} & \multicolumn{1}{c}{98.3} & \multicolumn{1}{c}{69.0} & \multicolumn{1}{c}{33.3} && \multicolumn{1}{c}{99.5} & \multicolumn{1}{c}{47.6} & \multicolumn{1}{c}{76.2} & \multicolumn{1}{c}{915.8} & \multicolumn{1}{c}{43.3}  \\ [0.2em]
\multicolumn{1}{c}{$\mathcal M_{G_2}$} & SQuAD2.0&\multicolumn{1}{c}{99.3} & \multicolumn{1}{c}{84.4} & \multicolumn{1}{c}{98.1} & \multicolumn{1}{c}{68.7} & \multicolumn{1}{c}{35.7} && \multicolumn{1}{c}{99.6} & \multicolumn{1}{c}{45.4} & \multicolumn{1}{c}{74.1} & \multicolumn{1}{c}{913.3} & \multicolumn{1}{c}{44.3} \\ [0.2em]
\multicolumn{1}{c}{$\mathcal M_{G_3}$} &CommonQA+NLI&\multicolumn{1}{c}{99.4} & \multicolumn{1}{c}{82.5} & \multicolumn{1}{c}{98.2} & \multicolumn{1}{c}{68.4} & \multicolumn{1}{c}{34.7} && \multicolumn{1}{c}{99.5} & \multicolumn{1}{c}{46.0} & \multicolumn{1}{c}{74.5} & \multicolumn{1}{c}{910.1} & \multicolumn{1}{c}{46.7} \\ [0.2em]
\multicolumn{1}{c}{$\mathcal M_{G_4}$} & NQ & \multicolumn{1}{c}{99.7} & \multicolumn{1}{c}{74.3} & \multicolumn{1}{c}{97.6} & \multicolumn{1}{c}{66.2} & \multicolumn{1}{c}{36.3}& & \multicolumn{1}{c}{99.6} & \multicolumn{1}{c}{35.1} & \multicolumn{1}{c}{64.2} & \multicolumn{1}{c}{884.4} & \multicolumn{1}{c}{50.8}  \\ [0.2em]\hline 
\end{tabular}}
\label{tbl:external_model_agnostic}
\end{table*}

\begin{table*}[!ht]
\centering
\caption{Attack Performance of \texttt{EMPRA} and baselines targeting the primary victim model $\mathcal{M}_V$ using the best-performing surrogate ($\mathcal{M}_{S_{\text{best}}}$) and generic ($\mathcal{M}_{G_{\text{best}}}$) NRMs over Easy-5 and Hard-5 target documents. $\downarrow$ indicates that lower perplexity and readability grade level is better. While multiple methods have attack success rates (ASR) close to 100\%, methods vary substantially in their ability to place the target document in the top-10 ($\%r\leq 10$), where they are more likely to be seen by the searcher.}
\begin{tabular}{c@{\hspace{1cm}}l@{\hspace{0.5cm}}cccccc}
\hline  \\[-1em]
 &  & \multicolumn{6}{c}{Easy-5}\\[0.2em]
 \cline{3-8} \\[-0.9em]
  Model  & Method & ASR & $\%r\leq 10$ & \multicolumn{1}{c}{$\%r\leq 50$} & \multicolumn{1}{c}{Boost} & \multicolumn{1}{c}{PPL$\downarrow$} & \multicolumn{1}{c}{Readability$\downarrow$} \\ \hline \\[-0.9em]
 & \texttt{Original} & \multicolumn{1}{c}{-} & \multicolumn{1}{c}{-} & \multicolumn{1}{c}{-} & \multicolumn{1}{c}{-} & 37.3 & 9.8 \\[0.2em]
- & \texttt{Query+} & 100.0 & 86.9 & 99.2 & 70.3 & 45.4 & 9.6 \\[0.2em]
 & \texttt{GPT-4} & 94.1 & 65.0 & 90.1 & 49.9 & 49.0 & 11.0 \\  \hline  \\[-0.9em]
 & \texttt{PRADA} & 77.9 & 3.5 & 46.2 & 23.2 & 94.4 & 9.9 \\[0.2em]
 & \texttt{Brittle-BERT} & 98.7 & 81.3 & 96.7 & 67.3 & 107.9 & 10.7 \\[0.2em]
$\mathcal{M}_{S_{\text{best}}}$ & \texttt{PAT} & 89.6 & 30.6 & 73.8 & 41.9 & 50.9 & 9.9 \\[0.2em]
 & \texttt{IDEM} & 99.7 & 87.4 & 99.0 & 70.3 & 36.4 & 9.4 \\[0.2em]
 & \texttt{EMPRA} & \textbf{99.9} & \textbf{95.6} & \textbf{99.8} & \textbf{72.5} & \textbf{34.4} & \textbf{9.2} \\ \hline  \\[-0.9em]
 & \texttt{PRADA} & 71.5 & 1.9 & 37.5 & 19.1 & 91.5 & 9.8 \\[0.2em]
 & \texttt{Brittle-BERT} & 90.0 & 43.4 & 80.1 & 46.2 & 117.7 & 11.0 \\[0.2em]
 $\mathcal{M}_{G_{\text{best}}}$ & \texttt{PAT} & 51.1 & 2.7 & 22.9 & 2.0 & 46.8 & 9.8 \\[0.2em]
 & \texttt{IDEM} & 98.8 & 65.3 & 93.8 & 61.9 & 37.7 & 9.4 \\[0.2em]
 & \texttt{EMPRA} & \textbf{99.7} &\textbf{74.3} & \textbf{97.6} & \textbf{66.2} & \textbf{36.3} & \textbf{9.2} \\
\hline

\\

\hline \\[-1em]
 &  & \multicolumn{6}{c}{Hard-5}\\[0.2em]
 \cline{3-8} \\[-0.9em]
  Model & Method & \multicolumn{1}{c}{ASR} & \multicolumn{1}{c}{$\%r\leq 10$} & \multicolumn{1}{c}{$\%r\leq50$} & \multicolumn{1}{c}{Boost} & \multicolumn{1}{c}{PPL$\downarrow$} & \multicolumn{1}{c}{Readibility$\downarrow$} \\ \hline \\[-0.9em]
 & \texttt{Original} & \multicolumn{1}{l}{-} & \multicolumn{1}{l}{-} & - & \multicolumn{1}{l}{-} & 50.5 & 9.0 \\[0.2em]
- & \texttt{Query+} & 100.0 & 47.8 & 78.3 & 955.1 & 67.5 & 9.0 \\[0.2em]
 & \texttt{GPT-4} & 99.3 & 28.7 & 59.4 & 873.8 & 58.7 & 10.2 \\ \hline  \\[-0.9em]
 & \texttt{PRADA} & 68.0 & 0.0 & 0.1 & 65.2 & 154.4 & 9.2 \\[0.2em]
 & \texttt{Brittle-BERT} & \textbf{100.0} & 61.5 & 85.9 & 965.5 & 152.5 & 10.1 \\[0.2em]
$\mathcal{M}_{S_{\text{best}}}$  & \texttt{PAT} & 98.0 & 6.2 & 20.1 & 589.1 & 71.4 & 9.2 \\[0.2em]
 & \texttt{IDEM} & 99.8 & 54.3 & 79.3 & 933.0 & 54.9 & 8.9 \\[0.2em]
& \texttt{EMPRA} & 99.9 & \textbf{64.9} & \textbf{87.0} & \textbf{948.4} & \textbf{47.1} & \textbf{8.8} \\ \hline  \\[-0.9em]
 & \texttt{PRADA} & 71.9 & 0.0 & 0.1 & 73.4 & 168.7 & 9.3 \\[0.2em]
 & \texttt{Brittle-BERT} & \textbf{99.9} & 17.7 & 47.6 & 845.2 & 156.8 & 10.3 \\[0.2em]
$\mathcal{M}_{G_{\text{best}}}$ & \texttt{PAT} & 79.0 & 0.0 & 0.7 & 92.9 & 64.2 & 9.0 \\[0.2em]
 & \texttt{IDEM} & 99.8 & 29.1 & 57.9 & 866.2 & 56.0 & 8.8 \\[0.2em]
 & \texttt{EMPRA} & 99.6 & \textbf{35.1} & \textbf{64.2} & \textbf{884.4} & \textbf{50.8} & \textbf{8.7}
\\\hline  
\end{tabular}
\label{table:surr_generic_models_results}
\end{table*}

\subsection{Surrogate-Agnostic Attack Performance Evaluation of \texttt{EMPRA}}

To answer \textbf{RQ1}, we evaluate the attack performance of \texttt{EMPRA} using six different surrogate and generic NRMs employed for generating adversarial documents targeting the victim model $\mathcal{M}_V$. The results, presented in Table \ref{tbl:external_model_agnostic}, report the impact of these models on the attack performance of \texttt{EMPRA}. As shown in the table, the first two models, $\mathcal M_{S_1}$ and $\mathcal M_{S_2}$, are surrogate NRMs trained on in-distribution data to mimic the behavior of the victim model. In contrast, the remaining models, $\mathcal M_{G_{1-4}}$, function as generic NRMs trained on out-of-distribution data, serving as relevance scoring models.

As shown in Table \ref{tbl:external_model_agnostic}, \texttt{EMPRA} consistently achieves high attack performance across all NRMs, highlighting its robustness and surrogate-agnostic adaptability to variations in relevance scoring. Specifically, it achieves an Attack Success Rate (ASR) exceeding 99\%, while successfully boosting over 97\% and 64\% of adversarial documents into the top-50 (\(\%r\leq50\)) on the Easy-5 and Hard-5 subsets, respectively. In terms of more stringent metrics such as boosted in top-10 (\(\%r\leq10\)) and how much the rank is boosted (Boost), \(\mathcal{M}_{S_1}\) achieves the highest \(\%r\leq10\) (\(95.6\%\)) and the largest rank boost (\(72.5\)) over Easy-5 subset, indicating that an in-distribution surrogate more effectively pushes adversarial documents to top ranks. Among the out-of-distribution generic NRMs \(\mathcal{M}_{G_1}\) (\(85.2\%\)) and \(\mathcal{M}_{G_2}\) (\(84.4\%\)) perform comparably well in terms of \(\%r\leq10\), illustrating that even OOD-based generic NRMs can cause substantial rank elevation.

For the more challenging Hard-5 subset, \(\mathcal{M}_{S_1}\) offers the highest \(\%r\leq10\) (\(64.9\)) and rank boost (\(948.4\)). Meanwhile, OOD-trained generic NRMs such as \(\mathcal{M}_{G_1}\) maintain competitive performance, achieving \(\%r\leq10\) values above 40\% with boosts over 900. Although ID-trained surrogates like \(\mathcal{M}_{S_1}\) provide a slight advantage in pushing adversarial documents to top-10 positions, OOD-trained generic models effectively undermine the victim model’s ranking consistency while producing fluent and coherent adversarial texts, as indicated by \(\mathcal{M}_{G_1}\)'s lowest perplexity (PPL) in both Easy-5 and Hard-5 subsets.

A critical observation is that the performance difference between surrogate NRMs and generic NRMs does not justify the significant computational cost and time required to train surrogate models in dynamic attack scenarios where victim models frequently change in architecture and relevance criteria. The surrogate-agnostic nature of \texttt{EMPRA} allows it to effectively leverage generic NRMs, eliminating the need for frequent retraining. This design choice enhances its practicality and establishes \texttt{EMPRA} as a versatile attack method adaptable to various ranking systems in real-world black-box attack settings.

\subsection{Attack Performance Comparison in ID and OOD Settings}

To answer \textbf{RQ2}, we evaluate the attack performance of our proposed method, \texttt{EMPRA}, in comparison with established state-of-the-art baselines targeting the primary victim model $\mathcal{M}_V$. The evaluation is conducted using the best-performing surrogate model (\(\mathcal{M}_{S_{\text{best}}}\)) to represent the in-distribution (ID) setting and the best-performing generic model (\(\mathcal{M}_{G_{\text{best}}}\)) to represent the out-of-distribution (OOD) setting for generating adversarial documents. These models, introduced in Section \ref{NRM_models}, are the best-performing surrogate and generic NRMs in terms of retrieval effectiveness, as previously mentioned and demonstrated by MRR@10 in Table \ref{table:mrr}. In addition, the underlying architecture of these models is based on the \texttt{BERT-base} model, which is also the foundation for several state-of-the-art baselines such as \texttt{Brittle-BERT}, thereby ensuring a fair and meaningful comparison. Table \ref{table:surr_generic_models_results} reports the attack performance of \texttt{EMPRA} and the baselines across these two models. Our analysis reveals several key findings:

\textbf{(i)} We observe that adversarial attacks consistently enhance document rankings across both surrogate and generic models and both sets of target documents. Notably, \texttt{EMPRA} demonstrates superior performance over all baselines across Easy-5 target documents and achieves higher attacking performance across almost all metrics on Hard-5 documents. Conversely, while \texttt{IDEM} generally performs well, it falls short of outperforming \texttt{Query+} in the majority of scenarios across both datasets. \texttt{GPT-4} occupies an intermediary position, surpassing trigger-based and word-level methods but lagging behind \texttt{Query+}, \texttt{IDEM}, and \texttt{EMPRA} in overall performance.

Lower boosted top-k values in Hard-5 target documents compared to Easy-5 ones are attributed to containing more irrelevant information relative to the query. Consequently, effective perturbations are required to increase exposure likelihood to users. \texttt{PRADA} and \texttt{PAT} exhibit limited effectiveness in boosting Hard-5 target documents within the top-10 or top-50. Conversely, \texttt{EMPRA} emerges as the most effective method, elevating nearly 65\% and 87\% of documents into the top-10 and top-50, respectively, using $\mathcal{M}_{S_{\text{best}}}$, while maintaining the lowest perplexity and readability grade level. This superiority over the best baseline, \texttt{IDEM}, amounts to a 19.52\% improvement in boosting top-10 and 9.70\% in boosting top-50 documents. This attack performance ($\%r\leq10$) improvement is important as boosting documents into the top-10 rankings is more valuable than other metrics due to the increased exposure to users. The superiority of \texttt{EMPRA} over other attack baselines, particularly \texttt{IDEM}, can be attributed to its ability to generate adversarial texts that not only maintain semantic proximity to the query but also with the top-ranked document and its sentences. This enables more effective adversarial texts that, when appended to the target document, significantly boost its ranking, as shown by metrics such as boosted top-k.

\textbf{(ii)} One of the main important aspects of an effective adversarial attack strategy is its robustness in attack performance against ID and OOD models. However, the performance of \texttt{PRADA}, \texttt{Brittle-BERT}, and \texttt{PAT} is heavily dependent on the surrogate models for adversarial text generation, resulting in lack of robustness across generic models trained on OOD data and a significant decrease in performance variability. For example, the attack performance of \texttt{Brittle-BERT} declines sharply when the external NRM is changed from $\mathcal{M}_{S_{\text{best}}}$ to $\mathcal{M}_{G_{\text{best}}}$, with the boosted top-10 value decreasing from 81.3\% to 33.2\% and 43.4\%, respectively. In contrast, \texttt{EMPRA} and \texttt{IDEM} exhibit greater stability across various external NRMs, as their adversarial document generation does not rely on any specific surrogate model characteristics, making them more adaptable to real-world attacking scenarios.

\textbf{(iii)} Another essential aspect in adversarial document generation is semantic fluency, measured through perplexity compared to baseline adversarial documents. Lower perplexity ensures that adversarial documents remain imperceptible and are less likely to be detected by detection models. Our findings reveal significant perplexity increases, particularly with \texttt{Brittle-BERT} and \texttt{PRADA}, notably evident in Hard-5 target documents. \texttt{Query+}, \texttt{GPT-4}, and \texttt{PAT} exhibit moderate levels of perplexity, striking a balance between complexity and fluency. However, adversarial documents generated by \texttt{Query+} are more easily detected and filtered due to just simply appending the query text to the document, thereby undermining the purpose of the attack. Notably, \texttt{EMPRA} surpasses \texttt{IDEM} in perplexity, particularly in Hard-5 target documents, claiming the top spot. This suggests \texttt{EMPRA} achieves superior attack performance while maintaining lower perplexity levels compared to both baselines and original documents in many cases. Additionally, \texttt{EMPRA} demonstrates the highest readability scores among the baselines by covering the lowest grade-level required for comprehension.

\begin{table*}[t]
\centering

\caption{Attack performance of adversarial documents using $\mathcal{M}_{S_{\text{best}}}$ against different victim NRMs.}

\begin{tabular}{llccccccccc}
\hline \\[-1em]
&  & \multicolumn{4}{c}{Easy-5} & & \multicolumn{4}{c}{Hard-5} \\ [0.2em] \cline{3-6} \cline{8-11}  \\[-0.9em]
\multicolumn{1}{l}{Victim  Model} & Method & ASR & $\%r\leq10$ & $\%r\leq50$ & Boost & & ASR & $\%r\leq10$ & $\%r\leq50$ & Boost \\ \hline \\[-0.9em]
& \texttt{PRADA} & \multicolumn{1}{c}{59.9} & \multicolumn{1}{c}{3.3} & \multicolumn{1}{c}{31.6} & \multicolumn{1}{c}{31.7} & & \multicolumn{1}{c}{35.3} & \multicolumn{1}{c}{0.0} & \multicolumn{1}{c}{0.1} & 4.9 \\ [0.2em] 
& \texttt{Brittle-BERT} & \multicolumn{1}{c}{98.5} & \multicolumn{1}{c}{83.6} & \multicolumn{1}{c}{95.8} & \multicolumn{1}{c}{132.3} & & \multicolumn{1}{c}{\textbf{99.9}} & \multicolumn{1}{c}{\textbf{73.5}} & \multicolumn{1}{c}{\textbf{88.0}} & \textbf{710.1} \\ [0.2em] 
\multicolumn{1}{l}{\texttt{ELECTRA}} & \texttt{PAT} & \multicolumn{1}{c}{88.6} & \multicolumn{1}{c}{29.0} & \multicolumn{1}{c}{66.6} & \multicolumn{1}{c}{89.5} & & \multicolumn{1}{c}{78.6} & \multicolumn{1}{c}{6.2} & \multicolumn{1}{c}{18.0} & 323.5 \\  [0.2em]
 & \texttt{IDEM} & \multicolumn{1}{c}{99.5} & \multicolumn{1}{c}{85.8} & \multicolumn{1}{c}{97.9} & \multicolumn{1}{c}{133.6} & & \multicolumn{1}{c}{98.2} & \multicolumn{1}{c}{56.3} & \multicolumn{1}{c}{75.9} & 667.4 \\  [0.2em]
 & \texttt{EMPRA} & \multicolumn{1}{c}{\textbf{99.8}} & \multicolumn{1}{c}{\textbf{92.2}} & \multicolumn{1}{c}{\textbf{99.3}} & \multicolumn{1}{c}{\textbf{136.1}} & & \multicolumn{1}{c}{97.7} & \multicolumn{1}{c}{66.8} & \multicolumn{1}{c}{84.0} & 685.9 \\  \hline  \\[-0.9em]
 
 & \texttt{PRADA} & \multicolumn{1}{c}{62.9} & \multicolumn{1}{c}{4.2} & \multicolumn{1}{c}{29.4} & \multicolumn{1}{c}{31.7} & & \multicolumn{1}{c}{57.7} & \multicolumn{1}{c}{0.0} & \multicolumn{1}{c}{0.4} & 27.4 \\  [0.2em] 
 & \texttt{Brittle-BERT} & \multicolumn{1}{c}{96.6} & \multicolumn{1}{c}{71.9} & \multicolumn{1}{c}{92.2} & \multicolumn{1}{c}{142.8} & & \multicolumn{1}{c}{\textbf{99.5}} & \multicolumn{1}{c}{57.3} & \multicolumn{1}{c}{78.4} & 731.9 \\  [0.2em]
 \texttt{DistilRoBERTa} & PAT & \multicolumn{1}{c}{85.7} & \multicolumn{1}{c}{25.6} & \multicolumn{1}{c}{61.0} & \multicolumn{1}{c}{90.1} & & \multicolumn{1}{c}{89.3} & \multicolumn{1}{c}{5.1} & \multicolumn{1}{c}{18.1} & 422.4 \\  [0.2em]
 & \texttt{IDEM} & \multicolumn{1}{c}{99.0} & \multicolumn{1}{c}{83.2} & \multicolumn{1}{c}{96.9} & \multicolumn{1}{c}{148.9} & & \multicolumn{1}{c}{98.9} & \multicolumn{1}{c}{57.3} & \multicolumn{1}{c}{78.3} & 724.3 \\ [0.2em]
 & \texttt{EMPRA} & \multicolumn{1}{c}{\textbf{99.6}} & \multicolumn{1}{c}{\textbf{89.6}} & \multicolumn{1}{c}{\textbf{98.6}} & \multicolumn{1}{c}{\textbf{152.3}} & & \multicolumn{1}{c}{97.8} & \multicolumn{1}{c}{\textbf{65.1}} & \multicolumn{1}{c}{\textbf{84.6}} & \textbf{735.5} \\ \hline 
\end{tabular}

\label{tbl:victim_agnostic_surrogate_NRM_result_table}
\end{table*}

\begin{table*}[!t]
\centering

\caption{Attack performance of adversarial documents using $\mathcal{M}_{G_{\text{best}}}$ against different victim NRMs.}

\begin{tabular}{llccccccccc}
\hline \\[-1em]
&  & \multicolumn{4}{c}{Easy-5} & & \multicolumn{4}{c}{Hard-5} \\ [0.2em] \cline{3-6} \cline{8-11}  \\[-0.9em]
\multicolumn{1}{l}{Victim  Model} & Method & ASR & $\%r\leq10$ & $\%r\leq50$ & Boost & & ASR & $\%r\leq10$ & $\%r\leq50$ & Boost \\ \hline \\[-0.9em]
 & \texttt{PRADA} & 52.8 & 2.2 & 27.1 & 25.7 & & 36.5 & 0.0 & 0.1 & 3.9 \\ [0.2em]
 & \texttt{Brittle-BERT} & 88.2 & 46.8 & 78.2 & 104.9 & & \textbf{97.6} & 20.7 & 47.3 & 576.2 \\ [0.2em]
\texttt{ELECTRA} & \texttt{PAT} & 57.0 & 3.0 & 26.0 & 18.6 & & 39.4 & 0.1 & 0.5 & -16.6 \\ [0.2em]
 & \texttt{IDEM} & 97.7 & 60.2 & 89.2 & 120.6 & & 96.1 & 27.2 & 51.8 & 574.4 \\ [0.2em]
 & \texttt{EMPRA} & \textbf{98.9} & \textbf{69.3} & \textbf{94.4} & \textbf{127.4} & & 96.2 & \textbf{35.8} & \textbf{60.0} & \textbf{605.9} \\  \hline  \\[-0.9em]
 & \texttt{PRADA} & 59.9 & 3.3 & 25.8 & 26.7 & & 58.3 & 0.0 & 0.2 & 32.4 \\ [0.2em]
 & \texttt{Brittle-BERT} & 86.9 & 39.3 & 73.4 & 112.2 & & \textbf{96.5} & 15.9 & 39.6 & 593.7 \\ [0.2em]
 \texttt{DistilRoBERTa}  & PAT & 52.9 & 2.5 & 22.6 & 13.4 & & 61.2 & 0.1 & 0.7 & 59.5 \\ [0.2em]
 & \texttt{IDEM} & 96.8 & 57.9 & 86.8 & 132.6 & & 97.1 & 27.2 & 52.9 & 633.5 \\ [0.2em]
& \texttt{EMPRA} & \textbf{98.1} & \textbf{64.5} & \textbf{91.4} & \textbf{140.1} & & 96.2 & \textbf{33.7} & \textbf{58.7} & \textbf{651.9} \\ \hline
\end{tabular}

\label{tbl:victim_agnostic_generic_NRM_result_table}
\end{table*}

\subsection{Attack Performance Evaluation Across Various Victim Models}
In real-world settings, attackers may not have detailed information about the target victim NRM used by search engines. In addition, due to continuous training, model updates, and potential replacement with different NRMs, the victim model changes frequently. As a result, an effective attacking method must perform reasonably across various victim models without the need for frequent retraining of surrogate models and regeneration of adversarial documents, which can be both costly and time-consuming. 

To answer \textbf{RQ3}, we adopt two different victim models and evaluate the performance of \texttt{EMPRA} and baseline attack methods using one surrogate NRM and one generic NRM on the same targeted Easy-5 and Hard-5 documents from \texttt{ms-marco-MiniLM-L-12-v2}. For this purpose, we compare the original rankings and the rankings after adversarial attacks when evaluated by the new victim model, with results presented in Tables \ref{tbl:victim_agnostic_surrogate_NRM_result_table} and \ref{tbl:victim_agnostic_generic_NRM_result_table}. Consistent with the previous section, we report how various victim NRMs rank adversarial documents generated by the best-performing surrogate NRM (\(\mathcal{M}_{S_{\text{best}}}\)) and the best-performing generic NRM (\(\mathcal{M}_{G_{\text{best}}}\)). These models, selected based on their retrieval effectiveness, represent the best-case (ID) and extreme-case (OOD) scenarios, respectively. This provides a comprehensive view of the attack method's robustness across varying victim model configurations.

Based on the results, \texttt{EMPRA} demonstrates the most robust performance in cross-victim NRM attacks compared to the baselines, exhibiting the lowest decrease ratio when transitioning from the ID surrogate model (\(\mathcal{M}_{S_{\text{best}}}\)) to the OOD generic model (\(\mathcal{M}_{G_{\text{best}}}\)), maintaining boosted top-50 rankings above 90\% across Easy-5 target documents and above 58\% across Hard-5 target documents. In contrast, \texttt{PRADA} and \texttt{PAT} exhibit the lowest attack performance due to their heavy reliance on the surrogate model for adversarial document generation, necessitating continuous surrogate model retraining for optimal performance, rendering them impractical for real-world applications. \texttt{IDEM} and \texttt{Brittle-BERT} occupy an intermediate position, displaying moderate attack performance. Notably, \texttt{Brittle-BERT} exhibits the highest attack performance when (\(\mathcal{M}_{S_{\text{best}}}\)) is used against \texttt{ELECTRA-base}; however, this performance diminishes by more than half when shifted to (\(\mathcal{M}_{G_{\text{best}}}\)), as evidenced in Table~\ref{table:surr_generic_models_results}, where its documents exhibit high perplexity, underscoring issues of quality and imperceptibility, as discussed in subsequent sections.
It is important to note that the average boost value exceeding 100 across Easy-5 target documents in Tables \ref{tbl:victim_agnostic_surrogate_NRM_result_table} and \ref{tbl:victim_agnostic_generic_NRM_result_table} occurs because the target documents were randomly sampled from rankings 51-100 of the primary victim model and may, in some cases, rank above 100 by the new victim models.

\begin{figure}[!t]

\centering
 \includegraphics[clip, trim= 0cm 0.25cm 0cm 0cm,scale=0.65]
 {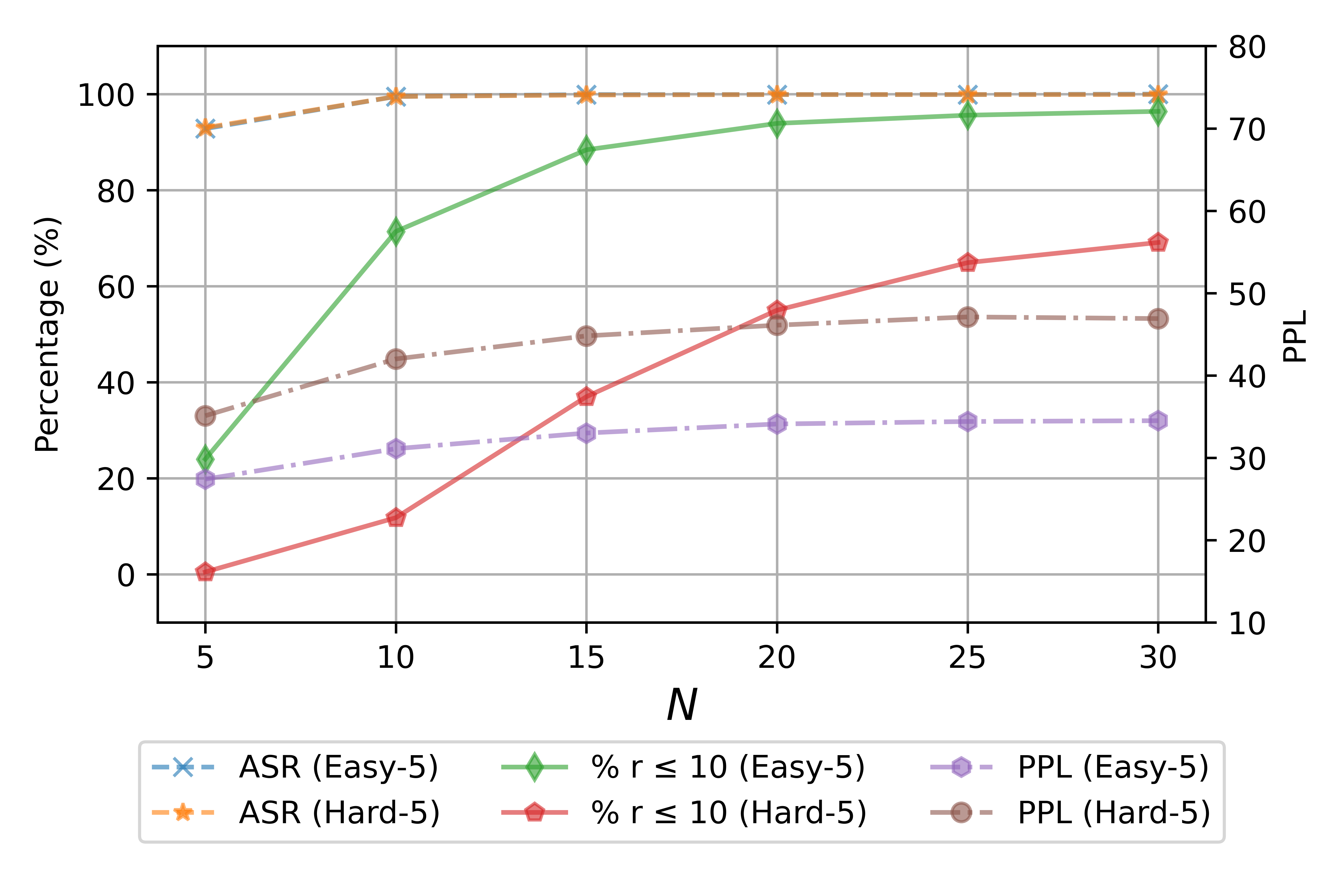}

\caption{Impact of the number of iterations.}

\label{fig:iterations_plot}

\end{figure}

\subsection{The Impact of Hyper-parameters} To answer \textbf{RQ4}, we evaluate \texttt{EMPRA} by exploring the impact of two hyper-parameters on its attack performance: 1) the number of iterations by the transporter function, and 2) the $\alpha$ interpolation coefficient, which balances relevance and coherence. The evaluation is conducted using the $\mathcal{M}_{S_{\text{best}}}$ model to analyze the performance of the proposed method under varying hyper-parameter settings.

Figure \ref{fig:iterations_plot} shows the impact of the number of iterations performed by the transporter function. We observe that as the number of iterations increases from 20 to 30, the improvement of attack performance by \texttt{EMPRA} becomes less substantial, particularly in comparison to the range of 5-20, especially noticeable with Easy-5 target documents. In terms of ASR, \texttt{EMPRA} can achieve comparable attack performance across both Easy-5 and Hard-5 target documents, indicating its capabilities of boosting both document sets.

\begin{figure}[t]

\centering
 \includegraphics[clip, trim= 0cm 0.75cm 0cm 0cm,scale=0.65]
 {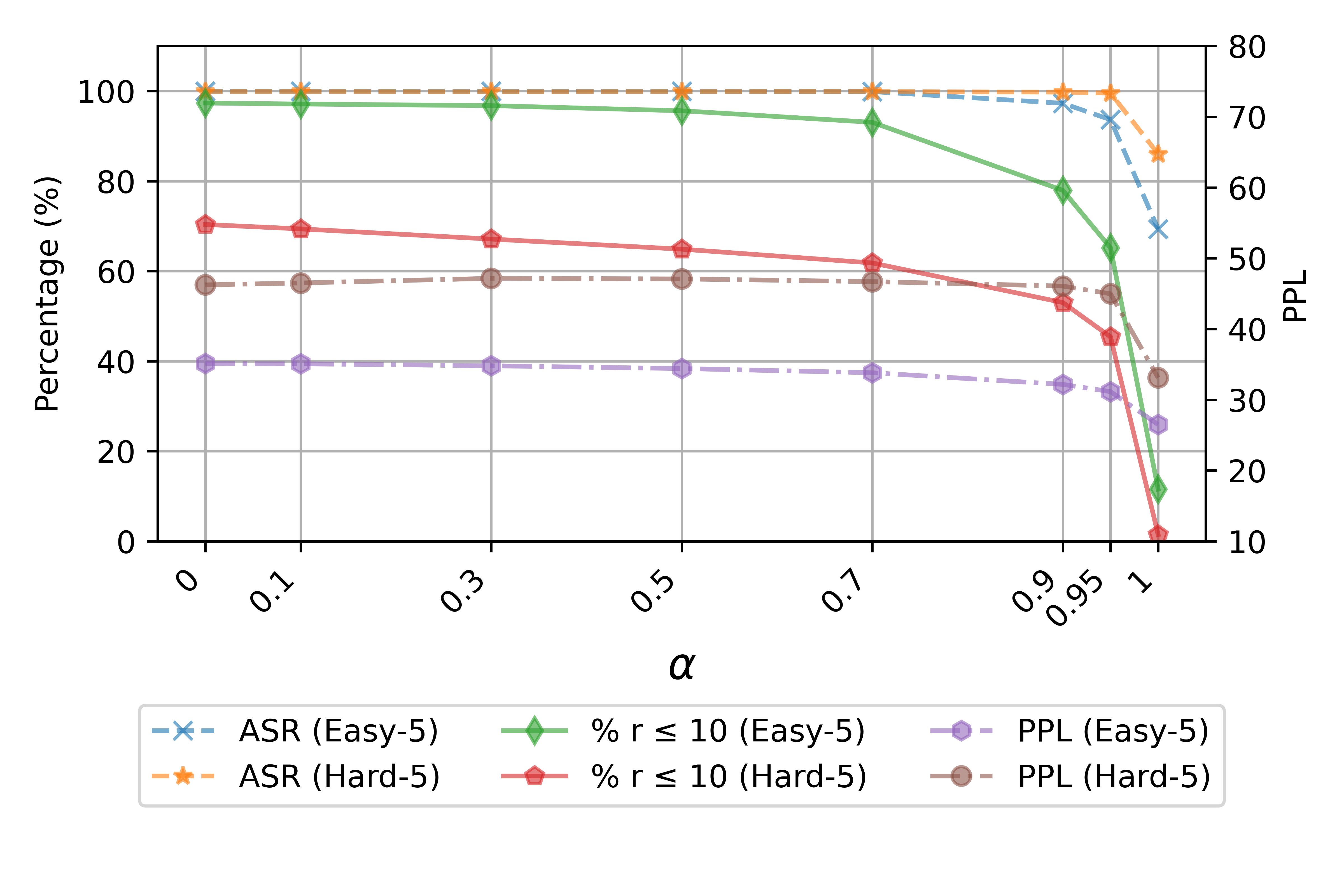}
 
\caption{Impact of the interpolation coefficient $\alpha$.}

\label{fig:alpha_plot}

\end{figure}

Moreover, Figure \ref{fig:alpha_plot} explores the impact of the interpolation coefficient $\alpha$ in Equation \ref{eq:interpolation} that balances between semantic coherence and query relevance. It is shown that when $\alpha$ falls within the range of 0-0.95, the attack performance remains consistently high, indicating that the adversarial sentences exhibit both strong attack capabilities and low perplexity. However, as the emphasis on coherency reaches its peak at $\alpha$ equal to 1, the attack performance begins to decrease, particularly in terms of boosted top-10.

\subsection{Quality and Naturalness Evaluation}\label{sec:naturalness}

In addition to evaluating attack performance, the quality, naturalness, linguistic acceptability, and imperceptibility of generated adversarial documents are important factors in maintaining reader confidence and achieving attack objectives. When reading the perturbed document the reader should not immediately suspect is has been manipulated. To assess these aspects and answer \textbf{RQ5}, we conduct an analysis to evaluate generated adversarial documents based on model and human evaluation metrics. Model-based metrics consist of text perplexity (PPL), grammar quality, and linguistic acceptability. Human-based evaluations metrics consist of imperceptibility and fluency, measured by human annotators. Due to limited space, we have provided the guidelines and details of our human annotation process in our repository. We evaluate the quality and naturalness of the `Mixture' target documents produced by each attacking method under \(\mathcal{M}_{S_{\text{best}}}\) model using model-based and human-based evaluation metrics. These results are compared with the overall attack performance ($\%r \leq 10$) in Table \ref{table:naturalness_evaluation}. A detailed explanation of the metrics follows.

\noindent\textbf{Model-Based Evaluation.} Perplexity, measured using the \texttt{GPT-2} model \cite{radford2019language}, serves as a proxy for fluency, with lower values indicating higher fluency. \texttt{EMPRA} achieves the lowest perplexity, indicative of its high fluency. To evaluate grammar quality, given the discontinuation of the Grammarly SDK as of January 10, 2024, we utilized the Grammarly website \cite{grammarly2023} to assess the overall quality score of each method's adversarial documents. Results indicate that documents generated by \texttt{EMPRA} closely match the quality of original documents in terms of grammar. \texttt{GPT-4} attains the highest quality scores, reflecting its proficiency in generating text, without even the errors that might be observed in the original document. However, its attack performance is considerably lower compared to \texttt{EMPRA}, particularly in boosting Hard-5 target documents.

\begin{table*}[t]
\centering
\caption{Trade-off between attack performance ($\%r \leq 10$) and the naturalness of adversarial documents generated by various attack methods. Naturalness is assessed using both model-based and human-based evaluation metrics. For ease of comparison, attack performance is taken from Table~\ref{table:surr_generic_models_results} (\(\mathcal{M}_{S_{\text{best}}}\)). \texttt{EMPRA} provides the best attack performance while maintaining among the best naturalness scores, often close to the original scores.}

\scalebox{0.85}{
\begin{tabular}{lcccccccccccc}
\hline \\ [-1em]
\multicolumn{1}{l}{Method} & \multicolumn{2}{c}{\begin{tabular}[c]{@{}c@{}}Attack Performance\\ ($\%r \leq 10$)\end{tabular}} & \multicolumn{5}{c}{Model-based Evaluation} & \multicolumn{4}{c}{Human-based Evaluation} \\ [0.2em] \cline{2-3} \cline{5-8} \cline{10-13}\\ [-0.9em]
\multicolumn{1}{l}{} & {Easy-5} & {Hard-5} &  & {PPL$\downarrow$} & {Grammar} & \begin{tabular}[c]{@{}c@{}}Acceptability\\ Score\end{tabular}  & \begin{tabular}[c]{@{}c@{}}Class.\\ Accuracy\end{tabular} & & {Impercept.} & {kappa} & {Fluency} & {kappa}  \\ [0.2em] \hline \\[-0.9em]
\texttt{EMPRA} & 95.6 & 64.9 & & 35.30 & 79.34 & 0.61 & 0.28 & & 0.59 & 0.48 & 3.42 & 0.01 \\[0.2em]
\texttt{IDEM} & 87.4 & 54.3 & & 39.27 & 80.44 & 0.64 & 0.28 & & 0.63 & 0.20 & 3.55 & 0.01 \\[0.2em]
\texttt{Query+} & 86.9 & 47.8 & & 45.03 & 74.91 & 0.46 & 0.53 & & 0.50 & 0.50 & 3.47 & 0.21 \\[0.2em]
\texttt{Brittle-BERT} & 81.3 & 61.5 & & 114.96 & 71.34 & 0.20 & 0.94 & & 0.47 & 0.01 & 3.36 & 0.09 \\[0.2em]
\texttt{GPT-4} & 65.0 & 28.7 & & 53.42 & 87.03 & 0.66 & 0.16 & & 0.64 & 0.26 & 3.72 & 0.35 \\[0.2em]
\texttt{PAT} & 30.6 & 6.24 & & 49.99 & 76.81 & 0.43 & 0.53 & & 0.55 & 0.31 & 3.52 & 0.03 \\[0.2em]
\texttt{PRADA} & 3.52 & 0.02 & & 126.24 & 51.72 & 0.40 & 0.69 & & 0.42 & 0.42 & 3.33 & 0.15 \\[0.2em]
\texttt{Original} & - & - & & 35.11 & 83.22 & 0.73 & 0.78 & & 0.50 & 0.29 & 3.59 & 0.02 \\[0.2em] \hline
\end{tabular}}

\label{table:naturalness_evaluation}
\end{table*}

To investigate the linguistic acceptability of the generated adversarial documents and explore whether they can be detected by trained Natural Language Processing (NLP) models, we employed the \texttt{RoBERTa-base} \footnote{\url{https://huggingface.co/textattack/roberta-base-CoLA}} classification model fine-tuned on the Corpus of Linguistic Acceptability (CoLA) \cite{warstadt2019neural} to specifically detect attacked texts. Using this model, we measure the linguistic acceptability scores of the original documents and their adversarial counterparts. In addition, the classification accuracy (denoted as 'Class. Accuracy' in the table) is also calculated to determine the accuracy of the model in detecting original documents vs adversarial documents correctly. 
The model correctly confirms that 78\% of the original documents have not been detected as adversarial documents. Moreover, the model's accuracy in classifying the adversarial documents generated by \texttt{EMPRA}, \texttt{IDEM}, and \texttt{GPT-4} is below 28\% demonstrating that over 70\% of these documents sufficiently resemble the original documents content and are linguistically acceptable, without containing any junk or garbage text. However, other baselines achieve a accuracy of more than 50\% having \texttt{Brittle-BERT} as the one with accuracy score of 94\%. This shows that despite its decent attack performance of \texttt{Brittle-BERT}'s adversarial documents, they can be easily and accurately detected using an NLP classification model, pointing that model has most likely added junk irrelevant trigger terms.

\noindent\textbf{Human-Based Evaluation.} Following prior studies~\cite{chen2023towards,wu2023prada,liu2022order}, to measure imperceptibility and fluency from a human perspective we recruited two annotators to assess the `Mixture' target documents for each attacking method. Annotators were tasked with determining whether the document content appeared to be manipulated (0) or not (1) given a query and target document, as well as assigning a fluency score ranging from 1 to 5. We calculated the average of annotator assessments for imperceptibility (denoted as 'Impercept.' in the table) and fluency over `Mixture' target documents for each attacking method and measured annotation consistency using the Kappa coefficient.

Our findings reveal that \texttt{PRADA}, \texttt{Brittle-BERT}, and \texttt{Query+} exhibit the lowest imperceptibility scores, while other attacking methods demonstrate higher imperceptibility compared to the original documents. This underscores their ability to maintain reader confidence and avoid raising red flags for the reader.

\noindent\textbf{Trade-off Between Attack Performance and Naturalness.} While evaluating the quality and naturalness of adversarial documents is essential, it is important to consider how these factors interact with attack performance. In this context, there should be a balance between achieving high attack performance and maintaining the naturalness and quality of the generated adversarial documents. For instance, although attacking methods like \texttt{Brittle-BERT} are effective in promoting the ranking of documents, they can easily be filtered out by linguistic acceptability models with the accuracy of 94\%, ruining the attack. 
In the trade-off between attack performance and naturalness, \texttt{EMPRA} excels by providing both high attack performance and high naturalness. While \texttt{GPT-4}, when prompted to generate an adversarial document, produces the most natural text according to linguistic measures and human assessment, \texttt{EMPRA} substantially outperforms it in terms of attack performance boosting the document among top-10 and increase the user exposure to adequately natural documents that maintain the attack objectives. At the same time, \texttt{EMPRA} outperforms other perturbation methods in this trade-off, emerging as a well-rounded approach that offers a balance between attack performance and naturalness.

\begin{table*}[t]
\centering
\caption{Adversarial documents generated using the \(\mathcal{M}_{S_{\text{best}}}\) model (except for \texttt{Query+} and \texttt{GPT-4}) by various attack methods targeting a document originally ranked at position 91 for the query “Can anyone take prenatal vitamins?”. The new ranking positions, perplexity scores, and linguistic acceptability of the generated documents are reported. Modified text is highlighted in \textcolor{LightRed}{red} for easy comparison, showcasing \texttt{EMPRA}'s superior balance of attack effectiveness, fluency, and linguistic acceptability.}
\label{tab:adv_examples}
\renewcommand{\arraystretch}{1.3} 
\scalebox{0.88}{
\begin{tabular}{l|p{9.75cm}|c|c|c}
\hline
\hline
\textbf{Method} & \textbf{Original or Adversarial Document} & \textbf{Rank$\downarrow$} & \textbf{PPL$\downarrow$} & \textbf{\begin{tabular}[c]{@{}c@{}}Linguistically\\ Acceptable\end{tabular}}\\ \hline

Original &
Always let your health care provider know what nutritional supplements you are taking. Prenatal vitamins consist of a variety of vitamins and minerals. During pregnancy, a woman's daily intake requirements for certain nutrients, such as folic acid (folate), calcium, and iron, will increase. & 
91 & 19.2 & Yes \\ \hline

\texttt{Query+} &
\textcolor{LightRed}{Can anyone take prenatal vitamins?} Always let your health care provider know what nutritional supplements you are taking. Prenatal vitamins consist of a variety of vitamins and minerals. During pregnancy, a woman's daily intake requirements for certain nutrients, such as folic acid (folate), calcium, and iron, will increase. &
1 & 20.4 & Yes \\  \hline

\texttt{GPT-4} &
\textcolor{LightRed}{Always consult your caregiver or medical specialist before starting any nutritional supplements, including those designed for prenatal care. Can individuals not expecting to conceive also consider prenatal vitamins? It’s a common inquiry. These vitamins and minerals blends, typically referred to as prenatal vitamins, are critical during gestation. During such a crucial timeline, a female body’s daily intake necessities for pivotal nutrients, such as folic acid (more commonly known by its synthetic form, folate), calcium, and iron, will see a notable escalation.} &
33 & 49.7 & Yes\\ \hline 

\texttt{PRADA} &
Always let your health care \textcolor{LightRed}{purveyor} know what nutritional supplements you are \textcolor{LightRed}{took}. Prenatal vitamins consist of a variety of vitamins and \textcolor{LightRed}{metallurgical}. During pregnancy, a woman's daily \textcolor{LightRed}{admitting} requirements for certain \textcolor{LightRed}{vitamin}, such as folic acid (folate), calcium, and iron, will increased. &
49 & 100.5 &  No\\ \hline

\texttt{Brittle-BERT} &
\textcolor{LightRed}{aanatnat anyone can...va taking 167 x <token> whether} Always let your health care provider know what nutritional supplements you are taking. Prenatal vitamins consist of a variety of vitamins and minerals. During pregnancy, a woman's daily intake requirements for certain nutrients, such as folic acid (folate), calcium, and iron, will increase. &
1 & 95.2 &  No \\ \hline

\texttt{PAT} &
\textcolor{LightRed}{no, if anyone could even take preca} Always let your health care provider know what nutritional supplements you are taking. Prenatal vitamins consist of a variety of vitamins and minerals. During pregnancy, a womanÃ¢Â€Â™s daily intake requirements for certain nutrients, such as folic acid (folate), calcium, and iron, will increase. &
1 & 33.8 &  No \\ \hline

\texttt{IDEM} &
\textcolor{LightRed}{Children, not pregnant mothers, cannot take prenatal vitamins.} Always let your health care provider know what nutritional supplements you are taking. Prenatal vitamins consist of a variety of vitamins and minerals. During pregnancy, a woman's daily intake requirements for certain nutrients, such as folic acid (folate), calcium, and iron, will increase. &
2 & 18.4 & Yes\\ \hline

\texttt{EMPRA} &
\textcolor{LightRed}{During pregnancy, anyone can take a prenatal vitamin (folic acid, iron, and calcium) to increase their daily requirements for these nutrients.} Always let your health care provider know what nutritional supplements you are taking. Prenatal vitamins consist of a variety of vitamins and minerals. During pregnancy, a woman’s daily intake requirements for certain nutrients, such as folic acid (folate), calcium, and iron, will increase. &
1 & 17.6 & Yes\\ \hline
\hline
\end{tabular}}
\end{table*}

\subsection{Adversarial Examples}\label{sec:adversarial_examples}

In order to have a better qualitative understanding of how each adversarial method perturbs the document, we present examples of adversarial documents generated by \texttt{EMPRA} and baseline methods, including \texttt{Query+}, \texttt{GPT-4}, \texttt{PRADA}, \texttt{Brittle-BERT}, and \texttt{IDEM} in Table~\ref{tab:adv_examples}. These documents were generated in response to the query: "Can anyone take prenatal vitamins?" and target a specific document that was originally ranked at position 91. All documents, except those generated by Query+ and LLM, were crafted using the \(\mathcal{M}_{S_{\text{best}}}\) surrogate model. The examples are evaluated based on their new ranking positions, perplexity scores, and whether they are deemed linguistically acceptable. 

As shown in the table, \texttt{Query+}, \texttt{Brittle-BERT}, \texttt{PAT}, \texttt{IDEM}, and \texttt{EMPRA} successfully promote the target document into the top-10 ranking positions for the given query. However, \texttt{Brittle-BERT} and \texttt{PAT} are flagged as unacceptable by the linguistic acceptability model (\texttt{RoBERTa-base}), indicating their adversarial modifications are easily detectable. Additionally, while \texttt{Brittle-BERT} and \texttt{PAT} achieve competitive ranking positions, they do so at the expense of significantly higher perplexity scores (45.03 and 114.96, respectively), undermining the fluency and naturalness of their generated documents. The reliance of \texttt{Query+} on inserting the exact query sequence into the document also makes it highly detectable, particularly if systems are designed to filter out documents containing such exact matches.

In contrast, adversarial modifications by \texttt{IDEM} and \texttt{EMPRA} are both subtle and impactful, as reflected in their lower perplexity scores and acceptance by the linguistic acceptability model. Notably, \texttt{EMPRA} outperforms \texttt{IDEM} by achieving a better ranking position (1st) for the target document and obtaining the lowest perplexity score among all methods (17.6). This indicates that \texttt{EMPRA} generates more fluent and natural text, making it a superior method for adversarial attacks. Its ability to craft high-quality, imperceptible modifications not only enhances ranking performance but also increases the likelihood of user exposure, underscoring its effectiveness in performing adversarial attacks.

\texttt{GPT-4} and \texttt{PRADA} are the only two methods that fail to promote the target document into the top-10 rankings, achieving positions 33 and 49, respectively. While \texttt{GPT-4} produces linguistically acceptable outputs, it suffers from the third-highest perplexity score and completely rewrites the target document. Furthermore, \texttt{GPT-4} struggles to optimize ranking performance effectively, highlighting its limitation in crafting highly targeted adversarial modifications. This makes it less suitable for adversarial scenarios where ranking promotion is a primary objective. \texttt{PRADA} demonstrates the poorest performance among all methods, achieving the lowest ranking boost and the highest perplexity score. Additionally, its output is not linguistically acceptable, with significant grammatical and semantic errors, such as "you are took" and "metallurgical." These issues make \texttt{PRADA} the least effective method in terms of both ranking performance and linguistic quality.

\section{Concluding Remarks}

In this paper, we introduced \texttt{EMPRA}, a novel method for executing adversarial attacks on black-box neural ranking models. \texttt{EMPRA} is a surrogate-agnostic method that operates independently of any specific surrogate model by utilizing two key components: the transporter and transformer functions. The transporter function adjusts sentence embeddings of the target document by shifting them closer to anchor texts, which include the target query, its top-ranked document, and the most similar sentence within that document. The transformer function then reconstructs these adjusted embeddings into coherent and fluent adversarial text. This dual-function design ensures that the generated adversarial texts can effectively manipulate the ranking of target documents while remaining imperceptible to human reviewers and automated systems when injected into the document.

\texttt{EMPRA}'s performance is demonstrated through its ability to significantly boost the rankings of target documents, particularly in challenging scenarios. In our evaluations on the MS MARCO V1 passage collection, EMPRA successfully re-ranked nearly 96\% of target documents from positions 51-100 into the top 10. Additionally, EMPRA elevated 65\% of hard target documents into the top 10 and 87\% into the top 50, showcasing its superiority over existing baselines. \texttt{EMPRA} stands out as a robust and adaptable attack method, highlighting the need for future research to develop defenses against such sophisticated adversarial technique.
\bibliographystyle{ACM-Reference-Format}
\bibliography{references}

\end{document}